\documentclass[a4paper, 11pt]{article}
\usepackage[T1]{fontenc}
\usepackage{amsmath,amssymb,amsfonts,amsthm,makeidx,graphicx}
\usepackage{txfonts}
\usepackage{helvet}
\usepackage{hyperref}

\newcommand{\bdec}{$\beta$-decay}
\newcommand{\bmdec}{$\beta^-$-decay}
\newcommand{\bspec}{$\beta$-spectrum}
\newcommand{\bspectra}{$\beta$-spectra}
\newcommand{\belec}{$\beta$-electron}
\newcommand{\krm}{$^\mathrm{83m}$Kr}
\newcommand{\ezero}{$E_0$}
\newcommand{\mnuetwo}{$m^2_\nu$}
\newcommand{\ccbc}{(license  \href{https://creativecommons.org/licenses/by/4.0/}{CC BY 4.0})}

\begin{document}


\title{KATRIN experiment}

\date{September 16, 2026}

\author{Guido Drexlin$^1$ and Christian Weinheimer$^2$\\[0.2cm]
$^1$ {\small Karlsruhe Institute of Technology, Institute of Experimental Particle Physics (ETP),}\\{\small Wolfgang-Gaede-Str. 1, 76131 Karlsruhe, Germany}\\
$^2$ {\small University of Münster, Institute for Nuclear Physics, Wilhelm-Klemm-Str. 9, 48149 Münster, Germany}\\
{\small Email: weinheimer@uni-muenster.de}
}

\maketitle

\centerline{\large \emph{Article for the Encyclopedia of Particle Physics, Elsevier}\vspace*{0.5cm}}

\begin{abstract} 
Since the discovery of neutrino oscillations, it is known that neutrinos have small but non-zero masses. The neutrino mass scale, which is of fundamental importance for cosmology, astrophysics and particle physics, can be measured directly from the kinematics of weak decays. The \underline{Ka}rlsruhe \underline{tri}tium \underline{n}eutrino experiment \emph{KATRIN}  measures the end point region of the tritium \bspec\ with unrivalled statistics and an unprecedented  precision. This world-leading direct neutrino mass search experiment is characterised by a windowless, gaseous molecular tritium source and a giant MAC-E filter-type spectrometer. The precision measurement of the tritium \bspec\ also allows the search for many other phenomena beyond the Standard Model of particle physics. The KATRIN experiment is about to reach its target sensitivity of the neutrino mass of less than 300 meV and will then turn its attention to the search for sterile keV neutrinos before the neutrino mass sensitivity is to be significantly increased once again by applying quantum read-out technology combined with an atomic tritium source with KATRIN++.
\end{abstract}


\begin{figure}[h]
	\centering
	 \includegraphics[height=8cm]{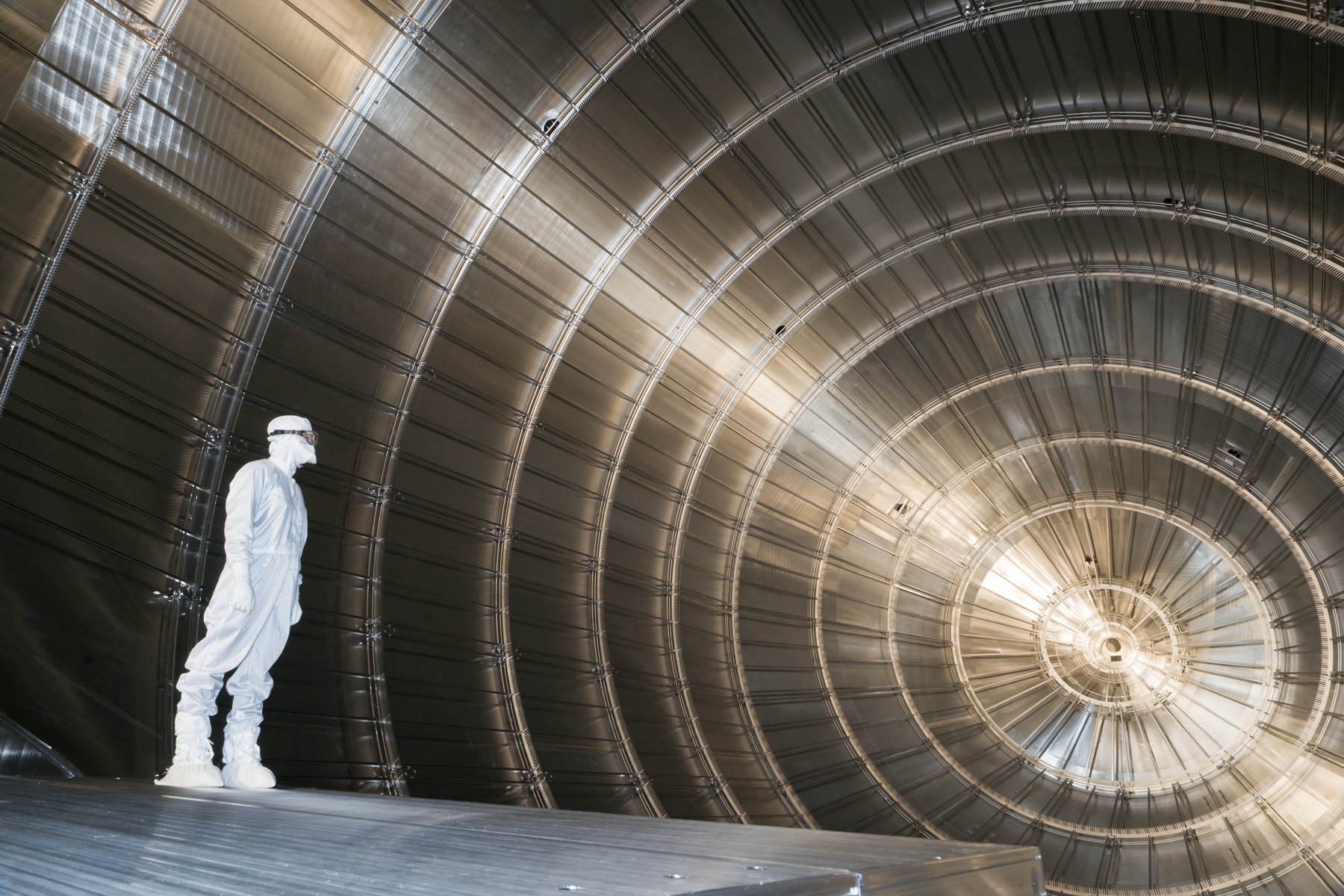 }
	\caption{A view inside the huge KATRIN spectrometer during installing the wire electrode system (photo and copyright: Michael Zacher). \label{fig:titlepage}}
	
\end{figure}


\paragraph{Nomenclature}MAC-E filter: spectrometer with magnetic adiabatic collimation combined with an electric retarding filter

\section*{Objectives}
\begin{itemize}
	\item   First, the reader will learn about Enrico Fermi's idea from 1934 on how the effective electron neutrino mass $m_\nu$ can be measured from the kinematics of a weak decay, e.g. from the shape of a \bspec\ near its endpoint. We will briefly relate this effective electron neutrino mass to the neutrino oscillation parameters and explain that $m_\nu$ is a kind of average of all neutrino mass eigenstate values that contribute to the electron neutrino. Ferm's idea is still valid today, and we will briefly present the history of experimental approaches since the late 1940s.
    \item In Section 2, the reader will learn about the KATRIN experiment in detail and learn that KATRIN's ultra-high statistics and low systematic uncertainties are due to the highly -luminous windowless gaseous tritium source, followed by a giant MAC-E filter-type spectrometer (see Fig. \ref{fig:titlepage}), as well as the many methods and precision tools used to determine the systematics.
    \item In Section 3, we explain how the measured data regarding the neutrino mass are analyzed from a convolution of the theoretical spectrum with the experimentally determined  \emph{response function}.
    \item Section 4 focuses on the physics investigated by KATRIN, namely the neutrino mass, but not only that:
    With its unprecedentedly precise data on the tritium \bspec\ below the end point, KATRIN can perform a whole range of further searches for physics beyond the Standard Model of particle physics, such as sterile neutrinos, generalised neutrino interactions, and cosmic neutrino background \dots
	\item   Section 5 presents other direct approaches to neutrino mass using the isotope tritium and the electron-capture isotope $^{163}$Ho and explains the connection to the search for neutrinoless double beta decay and cosmological analyses.
	\item   Section 6 provides an outlook on the next phases of KATRIN: In phase 2, the upgrade with the TRISTAN detector will allow searching for sterile keV neutrinos, an interesting candidate for warm dark matter, while the following phase KATRIN++ is intended to significantly improve the KATRIN experiment not in terms of size, but in terms of sensitivity in an intelligent way: Quantum technology is to be used to enable a differential measurement to drastically improve statistics, while an atomic tritium source is to be used to increase precision and reduce systematic uncertainties, with the first goal of achieving a neutrino mass sensitivity of 50\,meV to fully cover the scenario of inverted neutrino mass order.
    \item Section 7 contains the conclusions and concludes with the ambition to finally determine the neutrino mass in the laboratory.
\end{itemize}
\clearpage



\section{
Introduction -- history of the experiment 
}\label{history}
After Wolfgang Pauli had postulated in 1930 that that in \bdec , in addition to the emitted electron, a very light, electrically neutral and practically undetectable spin 1/2 particle must be emitted for the conservation of energy, momentum and angular momentum. Four years later, Enrico Fermi developed a theory of \bdec\ based on the assumption of a point-like current-current interaction and named the particle neutrino \cite{Fermi:1934hr}. This theory is still valid today as a low-energy approximation if it is extended by the principle of maximum parity violation, i.e. the vector operator 
describing the current is supplemented by an axial vector. 

In his important paper \cite{Fermi:1934hr}, Enrico Fermi also suggested how the mass of the neutrino could be determined: by measuring the \bspec\ near its end point. The \bspec\ $\dot N(E)$, the spectrum of the kinetic energy $E$ of the electrons emitted during \bdec , is essentially a pure phase space spectrum characterized by the product of momenta $p$ and relativistic energies $E_\mathrm{tot}$ of the two emitted leptons, electron and neutrino,  since the nucleus of mass number $A$, which is assumed to be very heavy, balances the momentum.
This spectrum does not require the introduction of an additional nuclear shape factor in case of an allowed or super-allowed \bdec s, but a correction function $F(A,Z+1)$, named after Fermi, which describes the electromagnetic interaction of the outgoing electron with the daughter nucleus $(A,Z+1)$, is required \cite{Otten:2008zz}:
\begin{equation} \label{eq:beta_spectrum_simple}
\dot N(E) \propto F(A,Z+1) \cdot \underbrace{\sqrt{E^2+2\cdot m \cdot E}}_{p_\mathrm{e}} 
\cdot \underbrace{(E+m)}_{E_\mathrm{tot,e}} \cdot \underbrace{(E_0-E)}_{E_\mathrm{tot,\nu}} \cdot \underbrace{\sqrt{(E_0-E)^2 - m^2_\nu}}_{p_\nu} \cdot \Theta(E_0-E-m_\nu)
\end{equation} 

As usual in the literature, the maximum kinetic energy of the electron for a vanishing neutrino mass $m_\nu=0$ is referred to as the endpoint energy \ezero , the electron mass is abbreviated  with $m$ and we use the calibration $c=1$.

It should be noted, that we will be speaking   through out this article of a neutrino and its mass $m_\nu$ here, even if the \bmdec\ of a nucleus  with charge number $Z$ involves an emitted electron antineutrino:
\begin{equation}
(A,Z) \rightarrow (A,Z+1)^+ + \mathrm{e}^- + \bar \nu_\mathrm{e}
\end{equation}

The factor $\sqrt{(E_0-E)^2 - m^2_\nu}$ yields the shape of the \bspec\ near its end point, from which Enrico Fermi suggested a method to determine the neutrino mass. 

\begin{figure}[h!]
    \centering
    \includegraphics[width=0.8\textwidth]{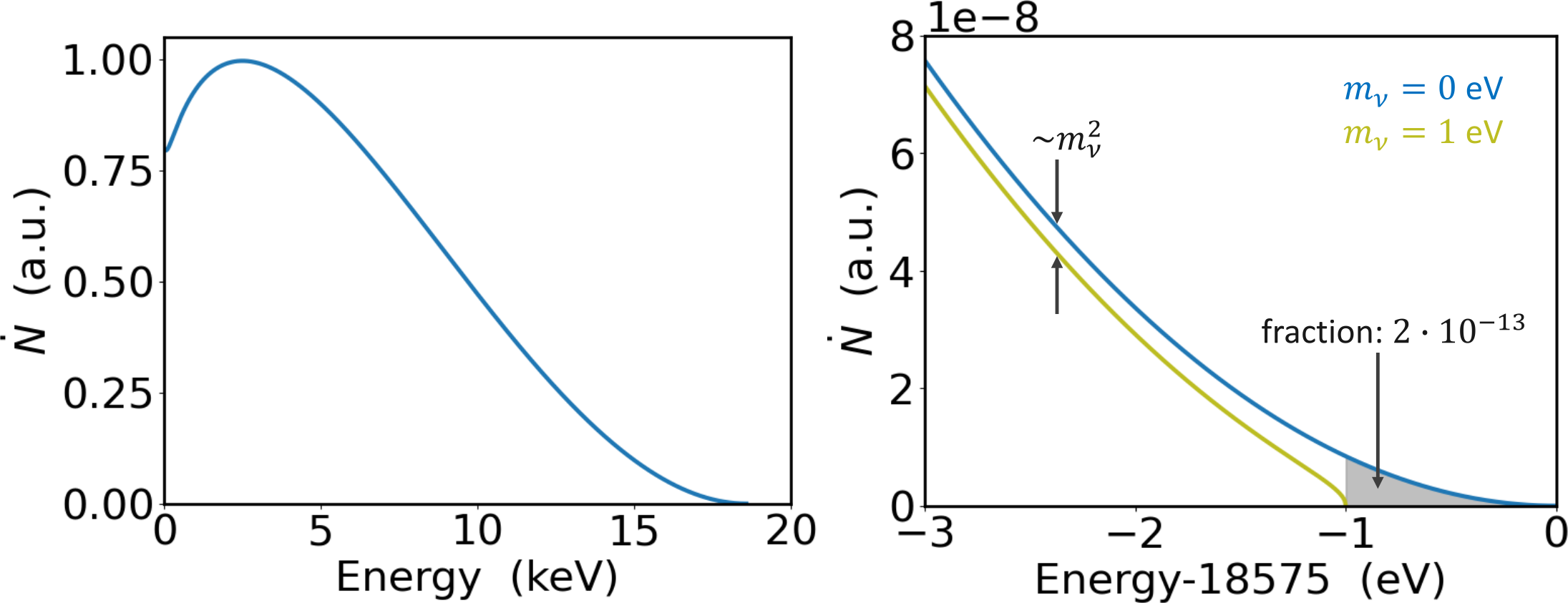}
    \caption{Beta spectrum of tritium (left) and its endpoint region for two assumed neutrino masses of $m_\nu=0$ and $m_\nu=1$\,eV (right), copyright Christian Weinheimer.
    \label{fig:tritium_beta_spectrum_at_endpoint}}
\end{figure}

Figure \ref{fig:tritium_beta_spectrum_at_endpoint} displays the \bspec\ of the superheavy hydrogen isotope tritium ($^3$H,T) on the left and a highly magnified section for hypothetical neutrino masses of $m_\nu=0$ and $m_\nu=1$ eV on the right. It becomes clear what the core of a direct experiment to determine the neutrino mass is. It must measure the shape of the \bspec\ in the endpoint region with high energy resolution. The energy resolution $\Delta E$ must be of a similar order of magnitude to the neutrino mass $m_\nu$ to be measured. Since the range over which the neutrino mass determines the shape is small and the count rate at the endpoint converges to zero, the \belec\ source must be exeptionally strong. Additionally, the endpoint energy \ezero\ should be small to maximize the interesting region of the \bspec . Tritium  fulfils all these requirements very well and, in addition, its \bdec\ is superallowed.

Since the late 1940s, physicists have been trying to measure the neutrino mass using the method proposed by Fermi \cite{Curran:1949pvk,Hanna:1949zab,Langer:1952til,Hamilton:1953ofv}. Tritium was almost always the $\beta$-isotope of choice. Until the 1980s, magnetic spectrometers were usually used, which, in addition to high energy resolution, also had a rather large solid angle acceptance in order to achieve the highest possible count rate in the endpoint region. In the 1970s, the energy resolution and sensitivity 
reached a level of precision that prompted
 Bergkvist decided to also take into account the atomic effects of a \bdec ing atom \cite{KarlErikBergkvist1971}. After \bdec , the daughter ion $(A,Z+1)^+$ is found in different electronic excitation states $V_j$ with an associated probability $W_j$, depending on the overlap of the original electron wave function in the atom $(A,Z)$ 
before decay with possible wave functions of this electron in the ion $(A,Z+1)^+$ 
afterward. The \bspec\ is therefore composed of many individual spectra, which must be summed up (or integrated in the continuum):
\begin{equation}
\dot N(E) \propto F(A,Z+1) \cdot p_\mathrm{e} \cdot (E+m) \cdot \sum_j W_j \cdot (E_0-E-V_j) \cdot \sqrt{(E_0-E-V_j)^2 - m^2_\nu} 
\cdot \Theta(E_0-E-V_j-m_\nu)
\label{eq:beta_spectrum}
\end{equation}

In the 1980s, a Russian group claimed \cite{Lyubimov:1980un,Boris:1987tq} that they had measured the neutrino mass of about 30\,eV from the \bspec\ of a tritiated hydrocarbon source  using a new magnetic spectrometer of Tretjakov type, featuring particularly high energy resolution and solid angle acceptance. This supposed discovery prompted many physicists worldwide to check the claim. With the tritium experiments initiated in the 1980s, particularly at Los Alamos \cite{Wilkerson:1987tr,Robertson:1991vn}, Zurich \cite{Fritschi:1986kd,Holzschuh:1992np} , Mainz \cite{Weinheimer:1993pd,Weinheimer:1999tn,Kraus:2004zw} and Troitsk \cite{Belesev:1995sb,Troitsk:2011cvm}, significant technical progress was made including the development of a windowless, gaseous molecular tritium source \cite{osti_6912423} and the development of a MAC-E filter-type spectrometer \cite{PKruit_1983,Lobashev:1985mu,Picard:1992kra}. The former allowed the \belec s  to leave the tritium source in a manner implying very little systematic uncertainty, while the latter is an integral electron spectrometer with extraordinary large energy resolution and solid angle acceptance.

After the discovery of atmospheric neutrino oscillation by the Super-Kamiokande experiment \cite{Super-Kamiokande:1998kpq}, the discovery of solar neutrino flavour transformation by the SNO experiment
\cite{SNO:2001kpb,SNO:2002tuh}, and the subsequent discovery of reactor antineutrino oscillation by the KamLAND experiment \cite{KamLAND:2002uet}, 
it became clear that neutrinos of flavours $\alpha$ ($\alpha = \rm{e}, \mu, \tau$) must mix with their neutrino mass eigenstates $\nu_i$ of mass $m_i$ ($i=1,2,3$) via a unitary $3 \times 3$ mixing matrix $U_{\mathrm{\alpha}i}$. Thus, the neutrinos must possess a mass, albeit a very small one. Since oscillation experiments, as a kind of interference experiment, can only determine the differences of neutrino mass squares $m^2_i-m^2_j$, but not the absolute neutrino masses, it became a high-priority for experimentalists to determine the neutrino mass using a direct method such as from the measurement of the endpoint spectrum of tritium \bdec . 
Therefore, the formula for the \bspec\ (\eqref{eq:beta_spectrum})  had to be extended by the three neutrino mass eigenvalues $m_i$ involved in the electron neutrino with its neutrino mixing matrix (\emph{Pontecorvo-Maki-Nakagawa-Sakata matrix}) elements $U_{\mathrm{e}i}$:
\begin{equation}
\dot N(E) \propto F(A,Z+1) \cdot p_\mathrm{e} \cdot (E+m) \cdot \sum_j W_j \cdot (E_0-E-V_j) \cdot \sum_i |U_{\mathrm{e}i}|^2 \cdot \sqrt{(E_0-E-V_j)^2 - m^2_i} \cdot \Theta(E_0-E-V_j-m_i)
\label{eq:beta_spectrum2}
\end{equation}

In practice, the individual neutrino mass eigenvalues $m_i$ cannot be resolved in direct neutrino mass experiments for the foreseeable future. Therefore, equation (\eqref{eq:beta_spectrum}) still holds when introducing an effective ``electron neutrino mass'': 
\begin{equation} \label{eq:electron_neutrino_mass}
  m_\nu = m(\nu_\mathrm{e}) := \sqrt{\sum_i |U_{\mathrm{e}i}|^2 \cdot m^2_i}
\end{equation}

In 2001, during a dedicated international workshop at Liebenzell Castle in the Black Forest, the KATRIN Collaboration was founded to develop, build and conduct a large experiment at the Karlsruhe Institute of Technology (KIT, formerly the Karlsruhe Research Center FZK). The $\beta$-source of the experiment was decided to be a windowless gaseous molecular tritium source, as demonstrated recently in Los Alamos, Livermoore and Troitsk, and a spectrometer of the MAC-E filter type, as successfully developed in Mainz and Troitsk. A condensed molecular tritium source, as developed at Mainz, was intended to serve as a backup solution. These technologies and methods were to be built on a significantly larger scale and further developed in the KATRIN experiment, along with many other essential new technologies and methods. The founding members of this international collaboration were therefore the scientists from the previous neutrino mass experiments in Los Alamos, Mainz and Troitsk, as well as the strong neutrino group of the just-concluded KARMEN neutrino experiment from Karlsruhe supplemented by other expert groups such as the conversion electron spectroscopy group of the Czech Academy of Sciences in Rez and a laser spectroscopy group from Swansea, Wales. The goal of KATRIN was to improve the sensitivity to the neutrino mass by an order of magnitude compared to the previously leading experiments in Mainz and Troitsk.
To achieve this goal would require an improvement of
 the sensitivity to the measured quantity or the neutrino mass square $m^2_\nu$ (see equation (\eqref{eq:beta_spectrum})) by two orders of magnitude. It took many years and 
many technological breakthroughs before before the KATRIN experiment could officially start with the first tritium spectrum collection in 2018. Another seven years later, the KATRIN experiment has delivered far more physics results than ever hoped for. Even though it has not yet found the neutrino mass, it is providing many exciting results that go far beyond the question of neutrino mass.

\begin{figure}[h!]
    \centering
    \includegraphics[width=0.8\textwidth]{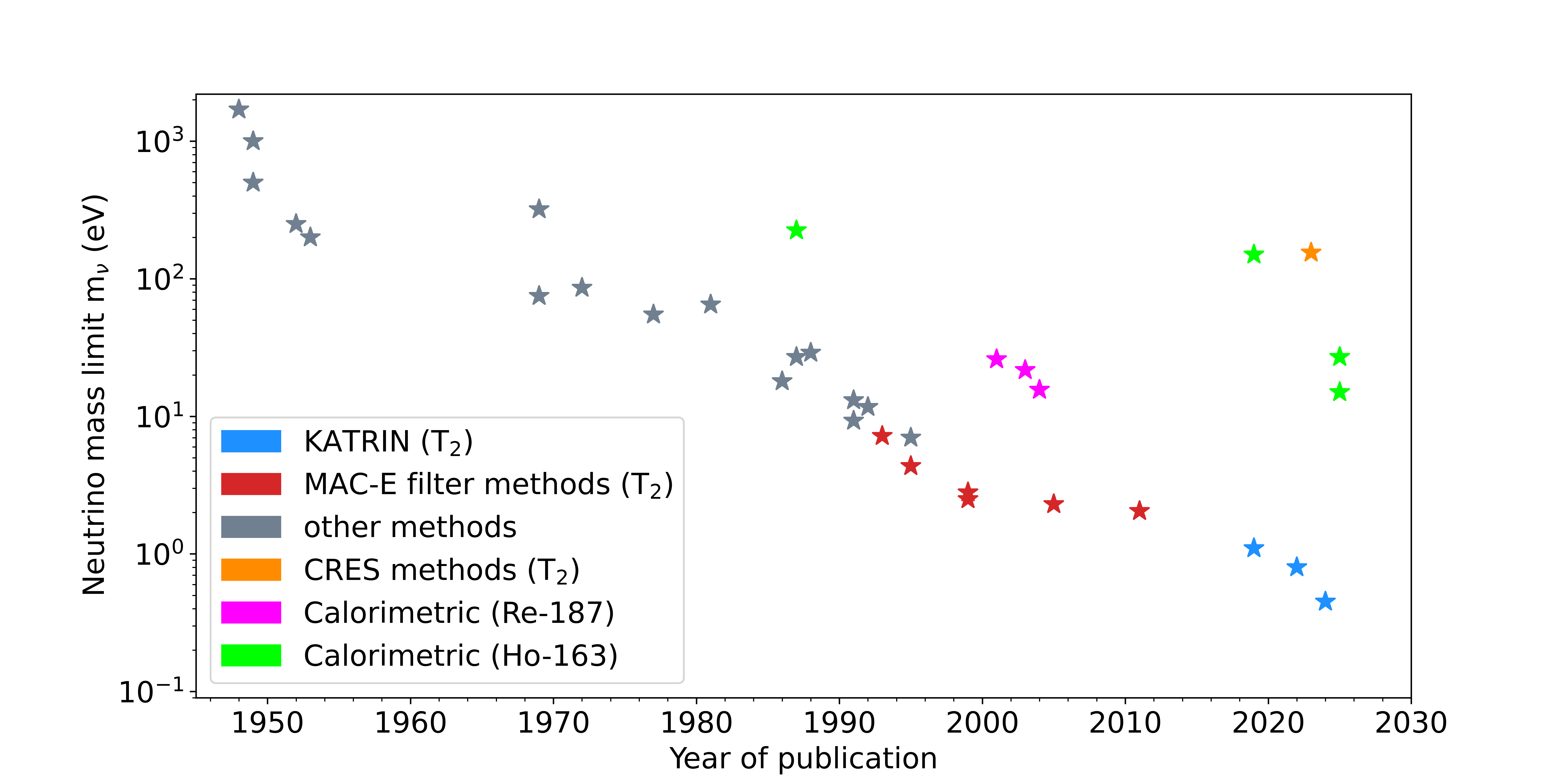}
    \caption{Neutrino mass limits from direct neutrino mass experiments over the last 80 years quoting the different technologies and isotopes used. The references on neutrino mass limits from tritium are \cite{Formaggio:2021nfz,Project8:2022hun,PhysRevD.110.030001,KATRIN:2024cdt}, the limits from $^{187}$Re are from references \cite{PhysRevD.110.030001,Camilleri:2008zz,Gatti:2001ty}, and those from $^{163}$Ho are from references \cite{Camilleri:2008zz, Velte:2019jvx,Alpert:2025tqq,lqkb-hylx} (courtesy and copyright: Magnus Schlösser).
    \label{fig:neutrino_mass_versus_time}}
\end{figure}

Figure \ref{fig:neutrino_mass_versus_time} shows the development of  direct neutrino mass searches from tritium \bdec\ and beyond over the last almost 80 years.

\clearpage


\section{Generic experiment description} \label{generic} 
\addtocounter{footnote}{-1}

The Karlsruhe Tritium Neutrino (\textsc{KATRIN}) experiment \cite{KATRIN:2021dfa} combines a high-intensity windowless gaseous tritium  \bdec\ source with a high-resolution electrostatic spectrometer of \textsc{MAC-E} filter type and a detector section. Both major components enable an unmatched spectroscopic information on the spectral shape of electrons close to the tritium endpoint $E_0 \approx18.6$\,keV. As shown in Fig. \ref{fig:neutrino_mass_versus_time}, and as outlined above, \textsc{KATRIN} is targeted to improve the experimental sensitivity on the effective electron neutrino mass $m_\nu$ down to 200 (300) meV. To achieve this huge gain in sensitivity, the key parameters of the experiment had to be designed such that after an overall net measuring time of 1000 days a statistical uncertainty of $0.018$\,eV$^2$ ($0.039$\,eV$^2$) and a corresponding systematic uncertainty of $0.017$\,eV$^2$ ($0.039$\,eV$^2$) on the experimental observable, the squared neutrino mass, $m_\nu^2$, could be reached. The numbers are from the \textsc{KATRIN} Design Report of 2004 while the numbers in brackets are the expectation after having collected most of the data. The main differences is a factor 10 higher background than anticipated (more details in the following sections).

\begin{figure}[h!]
    \centering
    \includegraphics[width=1\textwidth]{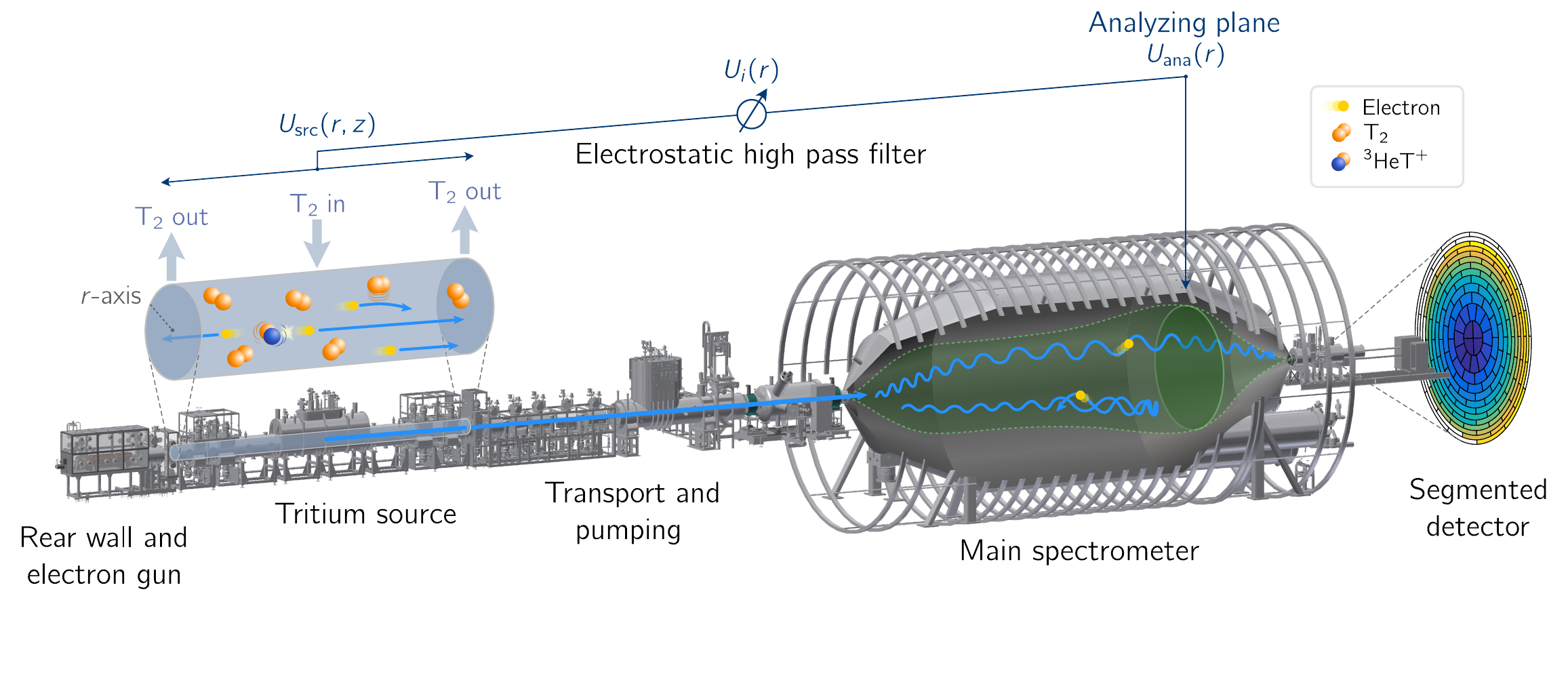}
    \caption{The KATRIN experiment with its major components from left to right: the calibration and monitoring system with its rear wall end electron gun, the windowless gaseous molecular tritium source WGTS, the electron transport section composed of a differential and a cryogenic pumping system, the small pre- and the large main spectrometer in the so-called SAP configuration, and the segmented detector 
    \cite{KATRIN:2024cdt_arXiv} \ccbc , see also \cite{KATRIN:2024cdt}.
     \label{fig:KATRIN_beamline}
     }
  
\end{figure}

\footnotetext[\thefootnote]{Readers may view, browse, and/or download material for temporary copying purposes only, provided these uses are for non-commercial personal purposes. Except as provided by law, this material may not be further reproduced, distributed, transmitted, modified, adapted, performed, displayed, published, or sold in whole or in part, without prior written permission from the publisher.}

Fig. \ref{fig:KATRIN_beamline} presents an overview of the \textsc{KATRIN} experiment.
The blue lines in Fig. \ref{fig:KATRIN_beamline} visualizes the journey of two \belec s starting by tritium \bdec\ in the windowless gaseous molecular tritium source  on the left. The \belec s spiral on cyclotron tracks along the magnetic field lines through the entire system from the tritium source to either up to the point of detection  of up to the point of reflection in the case that the retarding potential of the spectrometer will be higher than the electron's energy. To ensure a fully adiabatic transmission of the electrons the magnetic fields within the source and transport section are provided by superconducting solenoids.
Before outlining the key components of the \textsc{KATRIN} we fist explain the principle of the experimental design:

A windowless gaseous tritium source (WGTS), as developed by earlier neutrino mass experiments first at LANL and then in Troitsk and Livermoore, was chosen as the source for the \belec s. This principle ensures that the \belec s scatter only with tritium molecules within the WGTS, but not with other materials. This is a prerequisite for the intended highest accuracy in measuring the \bspec s with minimal systematic uncertainties. 
However, the windowless feature of the tritium source leads to further design decisions: Firstly, a windowless gaseous molecular tritium source cannot be implemented statically; instead, tritium gas is injected into the center of the WGTS and continuously pumped out at the ends.
This pumping process involves considerable effort at KATRIN. However, it is necessary in order to prevent tritium decay outside the WGTS, which would cause background events. The resulting column density of the WGTS must be kept constant at around one part per thousand and must also be known with this degree of accuracy. The former is achieved by precisely controlling the inlet pressure of the tritium injection and the temperature of the WGTS, while the latter is achieved by using a dedicated electron source further upstream, which allows the column density to be determined at regular intervals and absolutely by measuring the scattering  of electrons in the WGTS. 

\begin{figure}[h!]
    \centering
    \includegraphics[width=0.9\textwidth]{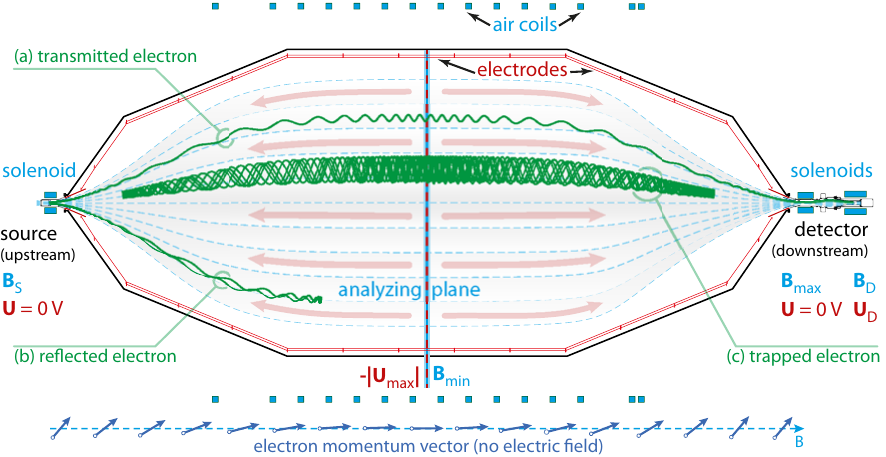}
    \caption{MAC-E filter at the example of the KATRIN main spectrometer in the normal, symmetric analysis plane configuration. A superconducting solenoid at the entrance and a superconducting solenoid at the exit together with the aircoil system form the magnetic field $B$ (field lines in grey and blue), whereas the electric potential at the spectrometer vessel (black) and the inner electrode system (red) form the retarding potential $U_\mathrm{max}$ ($q \vec E$ field lines in pink). The analysis plane is illustrated in the middle with the conditions  $B=B_\mathrm{min}$ and $-|U| = -|U_\mathrm{max}|$. Tracks of a transmitted, a reflected and a stored electron are illustrated in green. The third superconducting magnet housing the detector is exhibited as well. The diagram below illustrates the transmation of the momentum vector of the electrons according Eq. \eqref{eq:adiabatic_invariant} neglecting the electric retarding potential \cite{KATRIN:2021dfa} \ccbc .
\label{fig:KATRIN_MACE} }
\end{figure}

The second major design decision concerned the spectrometer. Due to its excellent energy resolution combined with large solid angle acceptance and moderate background rate, the KATRIN Collaboration opted for a spectrometer of the MAC-E filter type, as developed by the neutrino mass experiments in Mainz and Troitsk.
Such a MAC-E filter essentially consists of two (superconducting) solenoid coils, which generate a magnetic field similar to a magnetic bottle, and an electrode system between the two coils, which generates a retarding or analysis potential $qU$ by a negative countervoltage $U$ and the particle charge $q$ with $q=-e$ for electrons, see Fig.  \ref{fig:KATRIN_MACE}.
Such a retarding potential can be precisely operated for extended periods and measured, but has the disadvantage of only analyzing the \emph{longitudinal energy} $E_\parallel$, the energy component parallel to the magnetic (and by design, to the electric field lines), see Fig. \ref{fig:KATRIN_MACE}.
Converting most of the kinetic energy of the particle $E$  into longitudinal energy $E_\parallel$ prior to the energy analysis by the retarding potential -- by making use of the magnetic gradient force -- is the task of the inhomogeneous magnetic field,  which decreases by many orders of magnitude towards the center of the spectrometer. 

Under adiabatic conditions, the magnetic orbital moment 
\begin{equation}
\mu=\frac{E_\perp}{B} = \mathrm{const.} \label{eq:adiabatic_invariant}
\end{equation}
is a constant of motion (the equation is given here in a non-relativistic approximation). As a result, the energy component perpendicular to the magnetic field $E_{\perp}$ is converted into the energy component parallel to the magnetic field $E_{\parallel}$ in a decreasing magnetic field and vice versa in an increasing magnetic field.


Therefore, in the so-called \emph{analysis plane} (index \emph{min} for the minimal magnetic field) of the spectrometer, where the magnetic field is minimal and the retardating potential is maximal, almost all of the kinetic energy of the \belec s  is contained in the longitudinal component. For a \belec\ 
 that starts in the WGTS at an angle $\vartheta_\mathrm{s}$ relative to the magnetic field direction $\vec B_\mathrm{s}$ it reads:
\begin{equation}
E_{\parallel,\mathrm{min}} = E-E_{\perp,\mathrm{min}} = E \cdot (1- \sin^2\vartheta_\mathrm{s} \cdot B_\mathrm{min}/B_\mathrm{s})
\end{equation}
Thus, the transmission function $T(E,qU)$ of the spectrometer becomes an almost step-like function. 
The width of this transmission function $\Delta E$ is given by the maximum transverse energy of the \belec s in the analysis plane of the spectrometer over all starting angles $\vartheta_\mathrm{s}$.
In order to minimize the inelastic scattering of \belec s with tritium molecules in the WGTS, a maximum magnetic field $B_\mathrm{max}$ located at the downstream side of the spectrometer in \textsc{KATRIN} ensures, through the magnetic mirror effect, that only \belec s with a starting angle $\vartheta_\mathrm{s}$ smaller than the maximum starting angle of $\vartheta_\mathrm{s,max} = \arcsin(\sqrt{B_\mathrm{s}/B_\mathrm{max}})$ are transmitted to the detector.
Thus, the width of the transfer function is essentially defined by the ratio of the minimum and the maximum magnetic field times the energy:
\begin{equation}
\Delta E = E \cdot \frac{B_\mathrm{min}}{B_\mathrm{max}} \label{eq:DeltaE}
\end{equation}
The exact form of the transfer function for an isotropically emitting source is given by:
\begin{equation}
  T(E,qU) = \left\{ \begin{array}{ll} 
          0 & {\rm for~} E \leq qU\\
          1 - \sqrt{1 - \frac{E-qU}{E} \cdot \frac{B_\mathrm{s}}{B_\mathrm{min}}}
              & {\rm for~} qU < E < qU + \Delta E\\
           1 - \sqrt{1 - \frac{B_\mathrm{s}}{B_\mathrm{max}}}  & {\rm for~} E \geq qU + \Delta E 
     \end{array} \right. \label{eq:transmission_fcn}
\end{equation}

Since a spectrometer from the MAC-E filter accepts electrons up to an angle of $\vartheta_\mathrm{s,max}$, it can accept 
from sources such as the WGTS that emit isotropically in the full solid angle of $4 \pi$
a fraction of 
\begin{equation}
   \frac{\Delta \Omega}{4 \pi} = \frac{1}{2} \cdot \int_0^{\vartheta_\mathrm{s,max}} \sin \vartheta ' d\vartheta ' = \frac{1}{2} \cdot (1-\cos \vartheta_\mathrm{s,max}) = \frac{1}{2} \cdot \left( 1 - \sqrt{1 - \frac{B_\mathrm{s}}{B_\mathrm{max}}}~ \right)~,
\end{equation}
which, for the reference parameters of KATRIN, amounts to approximately 20\,\%  with excellent energy resolution, see equation \eqref{eq:DeltaE}.

However, there is also a disadvantage: with its magnetic configuration, a MAC-E filter can store charged particles in a magnetic bottle (see Fig.  \ref{fig:KATRIN_MACE}), which can then result in the ionization of residual gas atoms or molecules. The electrons released in this process typically have rather small initial energy, but follow the magnetic field lines. They can be accelerated either toward the spectrometer entrance or exit. If it is the spectrometer exit, these electrons are detected in the detector with practically the energy of the retarding potential $qU$ and cannot be distinguished energetically from signal electrons.

In the following, we outline the key features of the six sub-systems of \textsc{KATRIN} 
\begin{enumerate} 
\item {\bf Calibration and Monitoring system: }
Upstream of the WGTS there is the so-called calibration and monitoring system. It serves four different tasks: 
First, a rear wall that closes off the electron motion along the magnetic flux tube. Its gold plating ensures that the magnetic field lines of the flux tube all end on an electrically conductive plate with the same work function. An electrical potential can thus be applied to this rear wall to compensate for any differences between the work function of the rear wall and that of the WGTS beam tube in order to generate a homogeneous plasma potential in the WGTS. Furthermore, the measurement of the electric current to and from this rear wall can be used for plasma diagnosis. When the electrons hit the gold-plated rear wall, they generate characteristic gold X-ray lines, the intensity of which is measured by two lateral X-ray detectors. The intensity of the characteristic X-rays is one of the monitors of the intensity of the WGTS. A third task is provided by a UV light illuminating system which allows to inject photoelectrons from the rear wall into the WGTS. Before the launch of KATRIN it was not clear whether additional photoelectrons would have to be injected into the plasma of the WGTS to stabilize it temporally and spatially. After the launch of KATRIN, it quickly became clear that this UV illumination was not needed on a permanent basis.\\
Fourthly, the rear wall features a central hole with a diameter of 2 mm, through which electrons from a photoelectron source can be injected into the setup. The photocathode is the polished end of a gold-coated quartz fiber with a diameter of approximately 200\,$\mu$m. UV photons from a continuous laser-driven xenon light source sent through a monochromator or a pulsed UV laser can be used for illumination. An electrode system ensures that the electrons are first accelerated in a very strong electric field after they are generated. If this field is tilted relative to the magnetic field, electrons with defined starting angles can be generated depending on the electric field strength and tilt angle. The main tasks of this photoelectron source are, on the one hand, to determine the inelastic scattering of electrons in the WGTS, i.e., the energy loss function. On the other hand the electron gun allows to regularly determine the column density of the tritium gas in the WGTS, also by measuring the scattering rate. Superconducting deflection magnets in the inlet and outlet areas of the WGTS ensure that the beam from the photoelectron source can be guided through the WGTS at different radii in the magnetic flux tube for these important measurements.
\item {\bf Windowless gaseous molecular tritium source \textsc{WGTS}:} Tritium molecules (T$_2$) from a pressure-controlled buffer vessel, provided by the infrastructure of Tritium Laboratory Karlsruhe (TLK), are injected at the mid-point of the 10\,m long beam tube of the Windowless Gasaeous molecular Tritium source (WGTS), see Fig. \ref{fig:KATRIN_WGTS_principle} left. At both sides of the central WGTS beam tube and the down-stream (calibration and monitoring or rear system) and up-stream systems (DPS, CMS) the injected molecules are pumped back by a series of turbomolecular pumps with constant pumping speed (please see Fig. \ref{fig:KATRIN_WGTS_principle} right), resulting in a roughly triangular density profile with an integrated number of tritium molecules (labelled the column density $\rho d$) of $ \rho d \approx 4 \times 10^{17}/$cm$^2$ at operating temperatures ranging from initially 30\,K to 80\,K at present.
As the injected tritium gas has spurious amounts of the other hydrogen isotopologues it is continuously analyzed for its composition (T$_2$, HT, DT, D$_2$, H$_2$ und HD) using a laser Raman spectroscopy apparatus \cite{WOS:000570345700001}.
\begin{figure}[h!]
    \centering
    \includegraphics[width=0.9\textwidth]{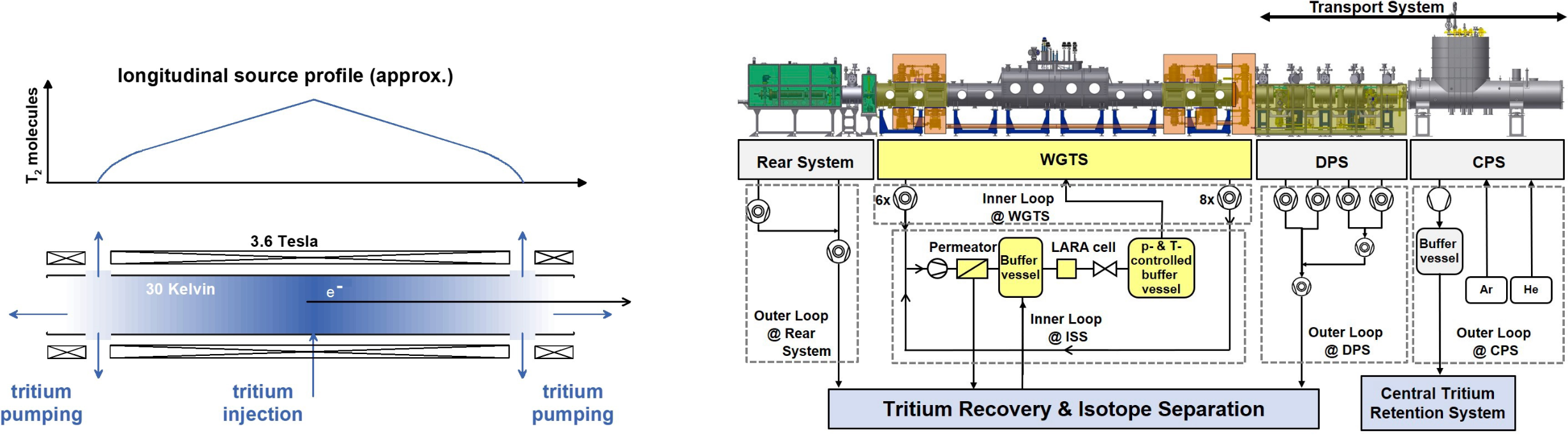}
    \caption{Principle of the WGTS consisting of the temperature-stabilized beam tube within a chain of superconducting magnets with the tritium injection in the middle and pumping out of tritium on its both ends (left), and the tritium loop system acting on the WGTS and its neighbouring systems (calibration and monitoring or rear system, DPS, CPS)  including pumping out, purification and injection back through a temperature- and pressure-stabilized vessel (right) \cite{KATRIN:2021dfa} \ccbc.}
    \label{fig:KATRIN_WGTS_principle}
\end{figure}

The high rate of $10^{11}$ \bdec s per second and the secondary ionization of tritium molecules in the WTGS by the \belec s generate a plasma in the WGTS. In order to control and stabilize the spatial and temporal stability of the plasma potential, various measures are taken, in particular with the calibration and monitoring system, as described above. In order to monitor their success and, in particular, to measure the plasma potential in situ where the \belec s are generated, small traces of the conversion electron emitter \krm\ can be injected into the WGTS together with the tritium gas. However, this requires the WGTS to be operated at 80 K  instead of the originally planned temperature of 30 K used in the first measurements in order to avoid the condensation of the krypton. This method has proven to be so powerful that all measurements since 2020 had been performed at this higher, now standard operating temperature of the WGTS. For spectroscopy measurements on \krm , where the background of \belec s from tritium interferes, pure \krm\ or this together with helium gas can also be injected. The latter mode allows higher and more uniform \krm\ rates from the WGTS.

\item {\bf Differential and cryogenic pumping section:} The WGTS turbomolecular pumps reduce the tritium pressure at both ends of the WGTS by two orders of magnitude already. However, this is far from sufficient to keep the spectrometer free of tritium molecules. Therefore, the WGTS is followed by a combination of differential pump and cryopump sections (see Fig. \ref{fig:KATRIN_DPS_CPS}). In addition to further molecular pumping in the differential pumping section, its five beam tubes are arranged in a zigzag line to prevent a direct line of sight and cause the tritium molecules to bounce off the beam tube walls, thereby drastically increasing their chance for pump off by a turbomolecular pump, resulting in a further reduction in tritium partial pressure by 5 orders of magnitude at the end of the differential pumping section. The subsequent cryogenic pumping section features a similar zigzag beam path at cryogenic temperatures. Its absolute temperature of $<5$ K and its coverage by a porous frozen argon layer provide very strong cryosorption, which reduces the tritium partial pressure by another 7 orders of magnitude, while the \belec s are guided without loss through the solenoids surrounding the beam tubes of the differential and cryogenic pumping sections.\\
Ring and dipole electrodes in the differential pumping section allow to block and drift out positively charged ions from the WGTS by the $\vec E \times \vec B$ force.
A side access between the last two beam tubes of the cryogenic pumping section serves the insertion of a \emph{forward beam monitor} to measure continuously the \belec s rate at the outer rim of the magnetic flux tube. A similar top access at the same point allows to bring in a calibration source: An isotropically emitting  electron source covering about 1\,\% of the magnetic flux tube can be positioned at any point of the flux tube. On its 25\,K cold graphite substrate the conversion electron emitter \krm\ can be frozen. With a pre-plating by stable krypton and continuous \krm\ feeding it provides an electron calibration source with longterm stability.
\begin{figure}[h!]
    \centering
    \includegraphics[width=0.5\textwidth]{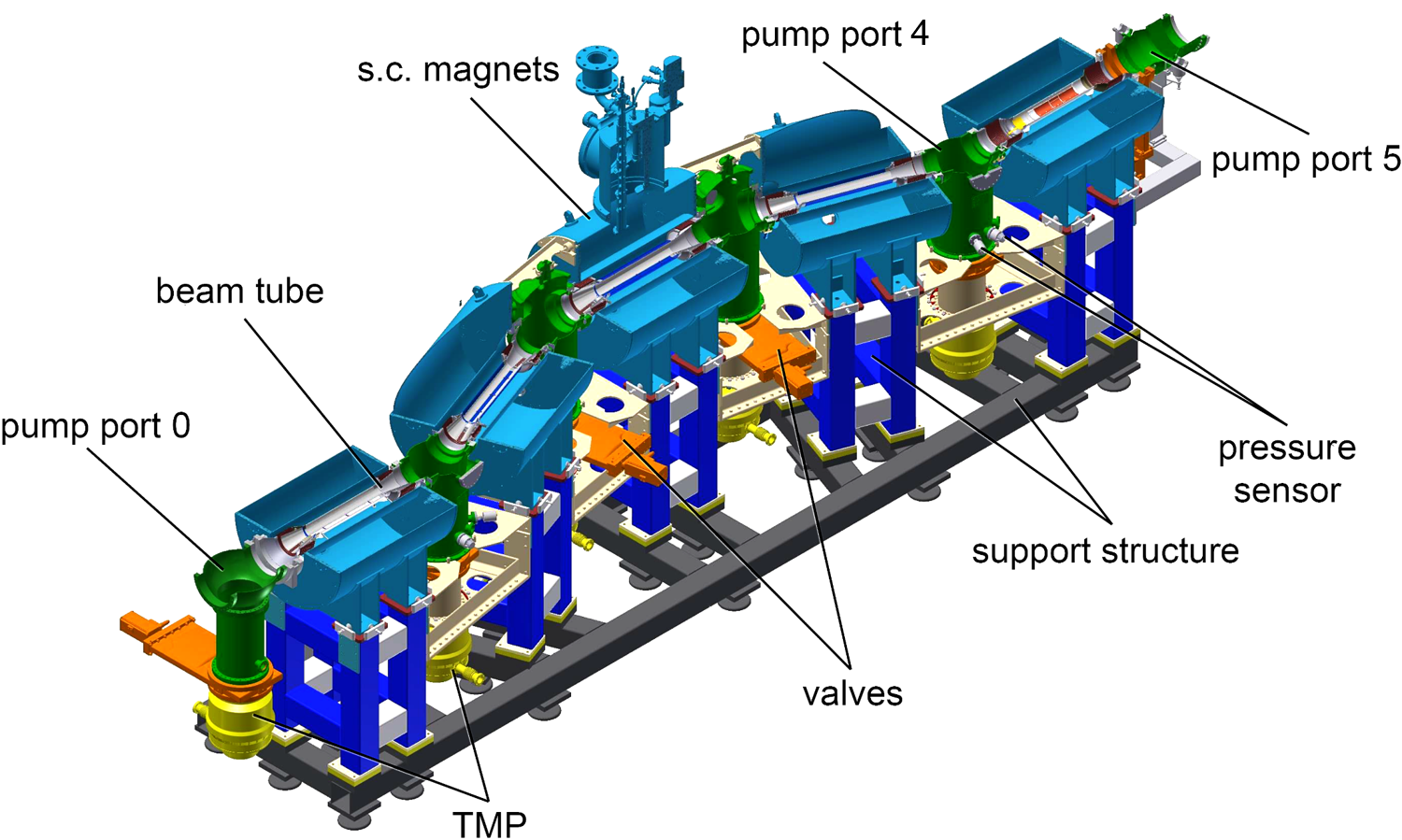}  \hspace*{0.2cm}   \includegraphics[width=0.45\textwidth]{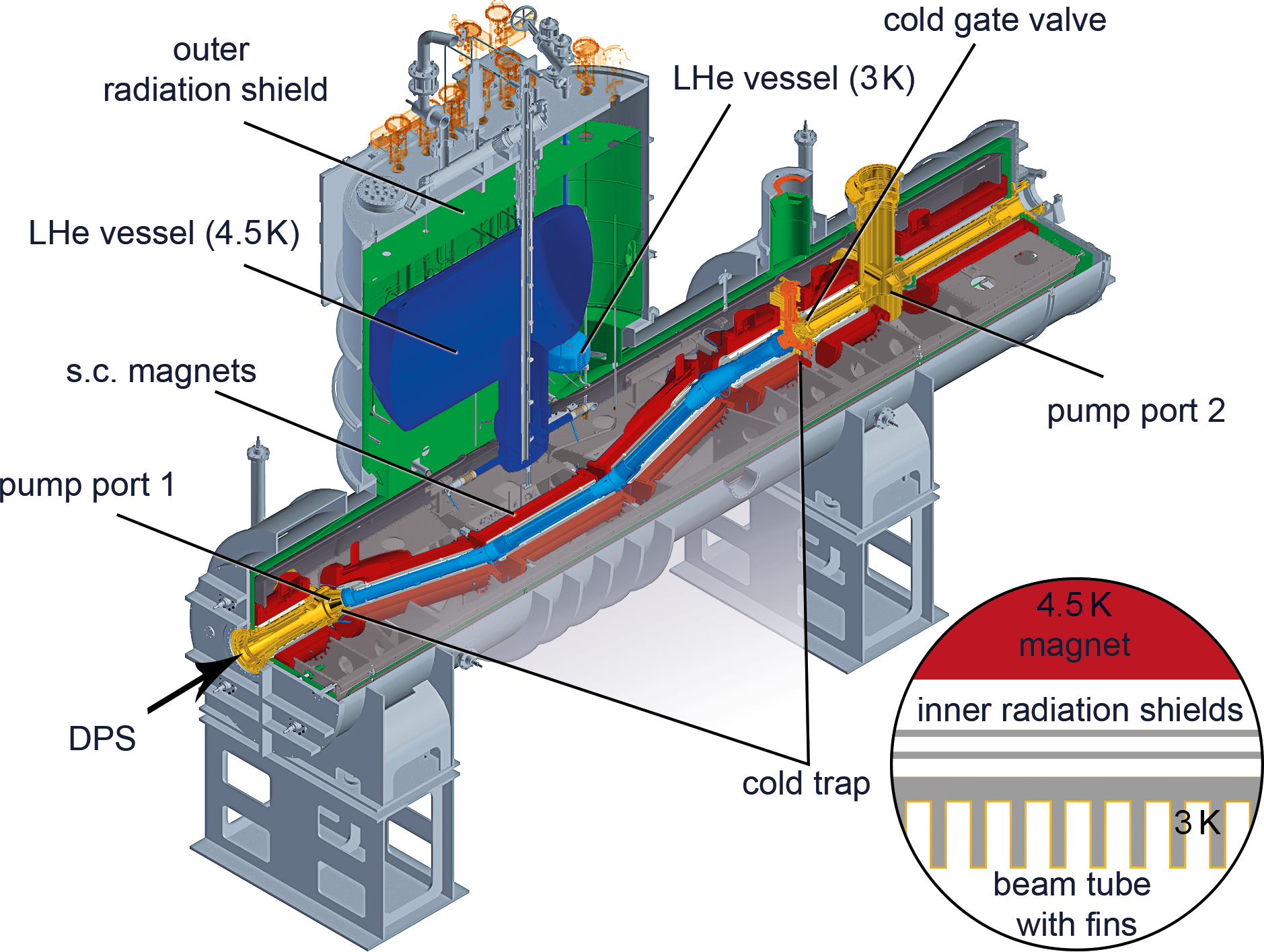}
    \caption{Electron transport and pumping system consisting of the differential pumping system and the cryogenic pumping system. Of the latter the $<5$\,K section is marked in blue and in the insert the structure of the baffles covered with argon frost are illustrated  \cite{KATRIN:2021dfa} \ccbc .
    \label{fig:KATRIN_DPS_CPS}}
\end{figure}

\item {\bf Pre-spectrometer:} The pre-spectrometer, a small MAC-E filter, was originally designed to allow only the high-energy portion of the \bspec\ to pass through to the large main spectrometer and reflect the rest. When the pre-spectrometer and main spectrometer are in operation, however, a Penning trap is created that can store electrons which leads to an increased background count rate. Although the Penning trap can regularly be emptied by a mechanical scraper, the time-dependent increase in the background rate between two emptying cycles compromises the intended precision and stability of the measurements, so that it is now only used with a very moderate retarding voltage of $-100$\,V. However, the pre-spectrometer also plays a very important role in the later series of measurements: It serves as an important barrier for tritium ions and neutral molecules originating from the WGTS in the direction of the large main spectrometer.
\item {\bf Main spectrometer:} The large or main spectrometer is impressive, first, because of its ultra-high vacuum conditions within its large dimensions, which is over 23 meters long and 10 meters in diameter, making it the largest ultra-high vacuum chamber in the world. The two superconducting solenoids 
in the region around the input and output ports, as well as
 the precision high voltage $U$ applied to the entire spectrometer, transform this arrangement into a MAC-E filter type electron spectrometer with very high resolution and solid angle acceptance. 
The inner, double-layer wire electrode system in the spectrometer (see Fig. \ref{fig:titlepage} and \ref{fig:KATRIN_MACE}) and the air coils around the spectrometer allow the electrical potential and magnetic field strength in the spectrometer to be precisely adjusted. This system makes it possible to maintain  adiabatic conditions over the entire electron path, resulting in the transformation of transverse energy into longitudinal energy, while also reducing the background rate from the spectrometer. Cosmic muons and radioactive trace amounts in the structural materials generate secondary electrons in the walls of the spectrometer, which are successfully prevented from penetrating the magnetic flux tube by the axial magnetic field and the counterpotential of the wires of the wire electrode system, which are at a more negative electrical potential.\\
The high voltage applied to the main spectrometer is monitored and actively stabilized by a wide-band system covering a frequency range of mHz to MHz to a ppm precision, i.e. $<20$\,mV at the typical operation voltage of -18.6\,kV. Moreover, the voltage can be absolutely calibrated to a voltage standard with a precision of one ppm as well. The precision voltage is given to the spectrometer vessel as base voltage and by low voltage supplies to the individual modules of wire electrode system.\\ 
The construction of the spectrometer vessel and the internal installation of the wire electrode system under normal laboratory air conditions has led to traces of the long-lived isotope $^{210}$Pb in the outer layers of the spectrometer walls. This can be traced back to the radioactive noble gas $^{222}$Rn, which emanates as a decay product of the trace element $^{238}$U, which is present practically everywhere, 
through the $\alpha$ decay of the isotope $^{214}$Bi from the $^{222}$Rn decay series. Even though this $^{210}$Pb activity is only about 1\,Bq/m$^2$ and the \bdec\ of $^{210}$Pb does not generate any background at the detector, it is the subsequent $\alpha$ decay of $^{210}$Po that leads to the sputtering of atoms from the stainless steel surface of the spectrometer wall due to its large recoil. These atoms can also be highly excited but neutral, enabling them to overcome the electrical and magnetic barriers of the wire electrode system and the axial magnetic field and fly into the spectrometer volume. If they are excited to very high main quantum numbers (e.g., $n>20$), they are referred to as Rydberg atoms, which can be ionized at room temperature solely by thermal radiation from the spectrometer walls. Other background electrons stem from neutral excited atoms where two electron orbitals in the atom that are excited in such a way that the state is auto-ionizing. In the KATRIN experiment, this process produces a background rate of over 1\,count per second detected in the detector and is proportional to the volume of the magnetic flux tube between the analysis plane and the detector. By baking the spectrometer and increasing the magnetic field in the spectrometer (at the expense of energy resolution according to equation \eqref{eq:DeltaE}), the background rate was reduced by a factor of 5. A further reduction by a factor of 2 was achieved by moving the analysis plane 5\,m closer to the detector through an optimized layout of
the magnetic fields of the air coils and the potentials at the wire electrodes (\emph{shifted analysis plane configuration}, SAP). The background rate is thus still a factor of 10 higher than originally intended, but since the experiment essentially determines the neutrino mass square from the shape of the \bspec\ near the endpoint \ezero , this only leads to a slight reduction in the original sensitivity to the neutrino mass from 200\,meV to below 300\,meV.

\item {\bf Detector: } 
Another superconducting solenoid contains the detector that counts the electrons transmitted by the main spectrometer. It is segmented into 148 pixels arranged in 12 concentric rings with 12 pixels each and another 4 pixels in the central circle. The detector, cooled to -30\,$^\circ$C, is housed in a lead shield and additionally equipped with a scintillation detector to veto cosmic muons for suppressing detector-related backgrounds. The detector and the charge-sensitive preamplifiers are located on a high-voltage platform of $-10$\.kV to sufficiently separate the electron signal from X-ray lines produeced by nearby materials. For that purpose a custom-made post-acceleration electrode connects the vacuum tube around the detector with that of the rest of the system.  The detector signals are digitized after the preamplifier and further processed by a fast data acquisition system that can process rates of up to 1\,MHz with virtually no dead time. Fig. \ref{fig:KATRIN_FPD} shows the detector system including its calibration units.

\begin{figure}[h!]
    \centering
    \includegraphics[width=0.7\textwidth]{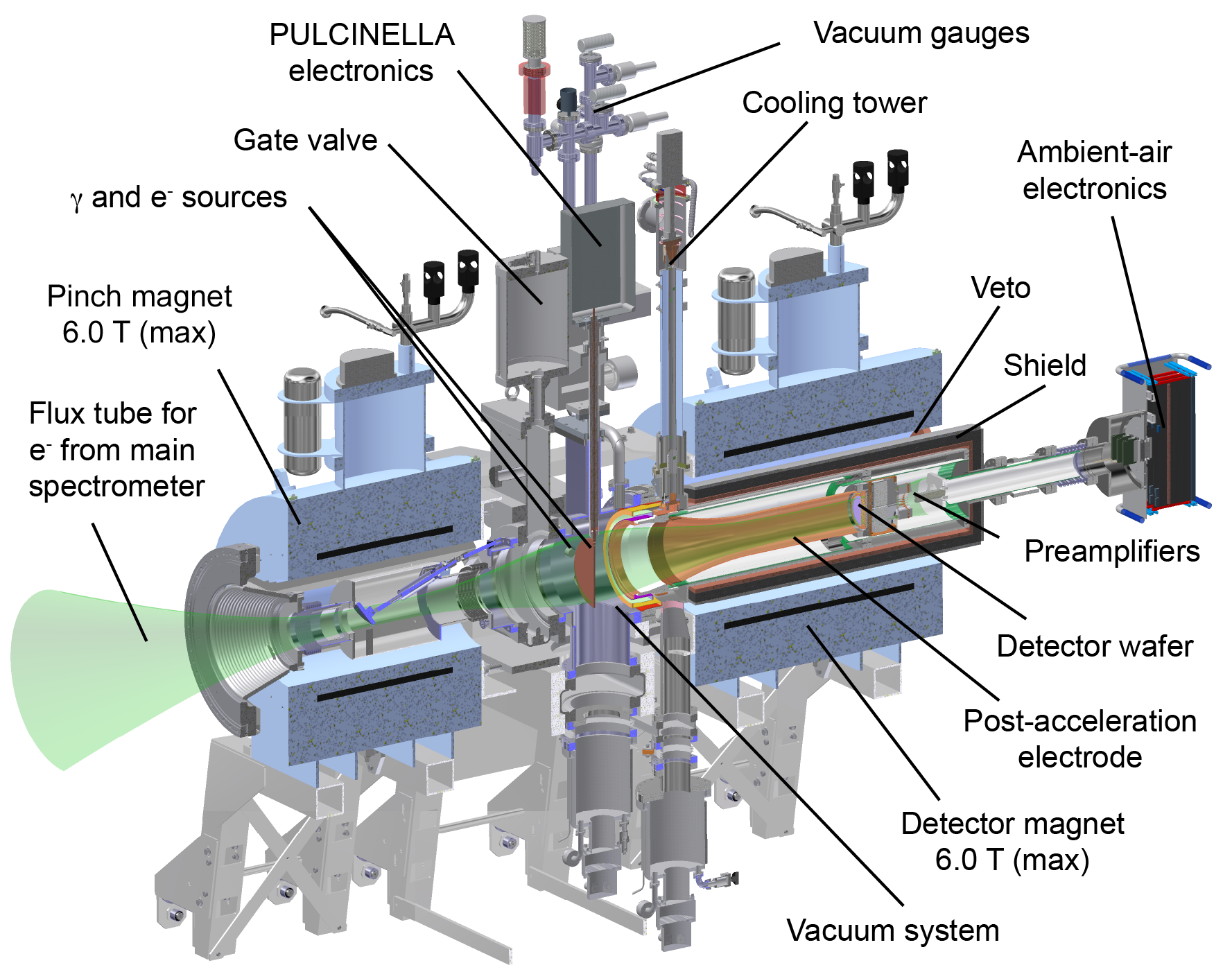}
\caption{Detector system with its segmented silicon detector housed in a dedicated detector magnet. For calibration of the detector with gamma- and X-rays a $^{241}$Am source can be brought into the beam tube. Calibrating the detector with photoelectrons is done by inserting the PULCINELLA disc into the beam tube, putting it on a negative high voltage and radiating it with ultraviolett light \cite{KATRIN:2021dfa} \ccbc.
 \label{fig:KATRIN_FPD}}
\end{figure}

\end{enumerate}

\clearpage

\section{From data to physics}\label{data}

\addtocounter{footnote}{-2}

\begin{figure}[h!]
    \centering
    \includegraphics[width=0.8\textwidth]{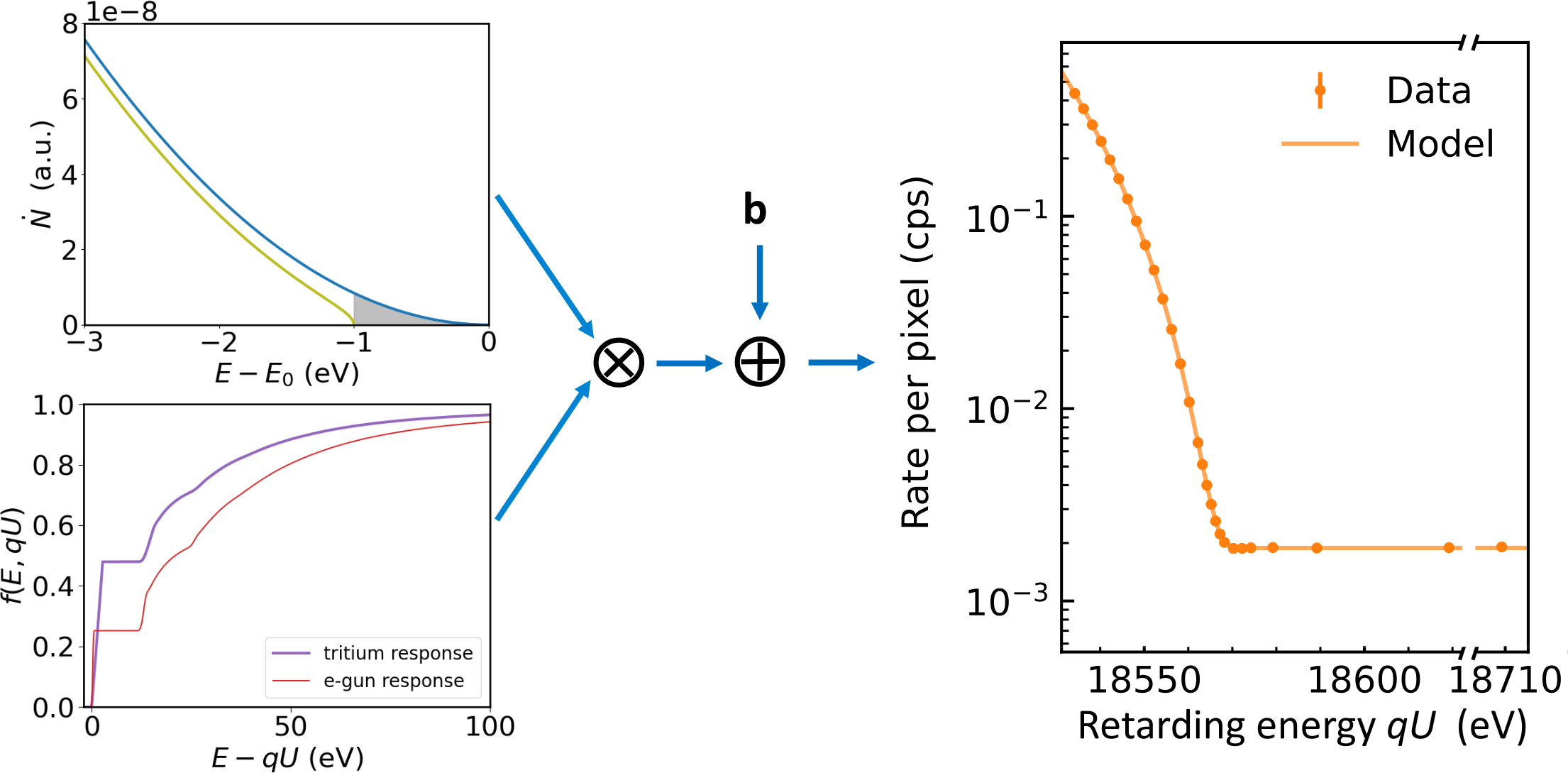}
    \caption{Convolution of the theoretical \bspec\ (top left) with the (tritium) response function (bottom left) and adding a background $b$ describes the experimental measured data (right). Please note the difference between the e-gun response and the tritium response, because electrons from the electron photoelectron source in the calibration and monitoring system transverse the full WGTS, where \belec s from tritium decays within the WGTS in average transverse only half of the WGTS, copyright Christian Weinheimer.
 \label{fig:beta_spectrum_response_convolution}}
\end{figure}

At KATRIN, the tritium \bspec\ is determined by measuring the \belec\ rate transmitted by the spectrometer at specific retarding energies, which are defined by the high voltage set points on the spectrometer. Figure \ref{fig:beta_spectrum_response_convolution} shows the count rate as a function of the spectrometer retarding energy for a series of measurements on the right. At each of the approx. 35 retarding energies, repeated measurements are carried out between 30\,sec  and 21\,min.

Due to the very sharp, integrating transmission function, one would expect, as a simplified model function describing the measurement data, the integral of the tritium spectrum above the respective retarding energies, or more precisely, the convolution of the \bspec\ with the transmission function at the respective retarding energies, see Eq. \eqref{eq:transmission_fcn}. However, some of the \belec s scatter inelastically one or more times in the tritium source, so that the convolution with the transmission function can only describe the non-scattered \belec s. Therefore, to describe the measured count rate $R(qU)$ the \bspec\ is convolved with the so-called \emph{response function} $f(E,qU)$ instead of the spectrometer's transmission function $T(E,qU)$ from Eq. \eqref{eq:transmission_fcn}
, which also takes into account single and multiple scattering. In addition, an energy-independent flat background rate $b$ must be added, which originates from electrons produced in the spectrometer or is measured at the detector due to gamma activity, X-rays and muons:
\begin{equation}
  R(qU) = A_\mathrm{s} \cdot N_\mathrm{T} \int_{qU}^{E_0} \dot N(E) \cdot f(E,qU) ~ dE + b \label{eq:fit_fcn_beta_spec}
\end{equation} 
Here $A_\mathrm{s}$ is a normalisation constant whereas $N_\mathrm{T}$ is the number of tritium atoms in the WGTS.
\begin{figure}[h!]
\centering
\includegraphics[width=0.6\textwidth]{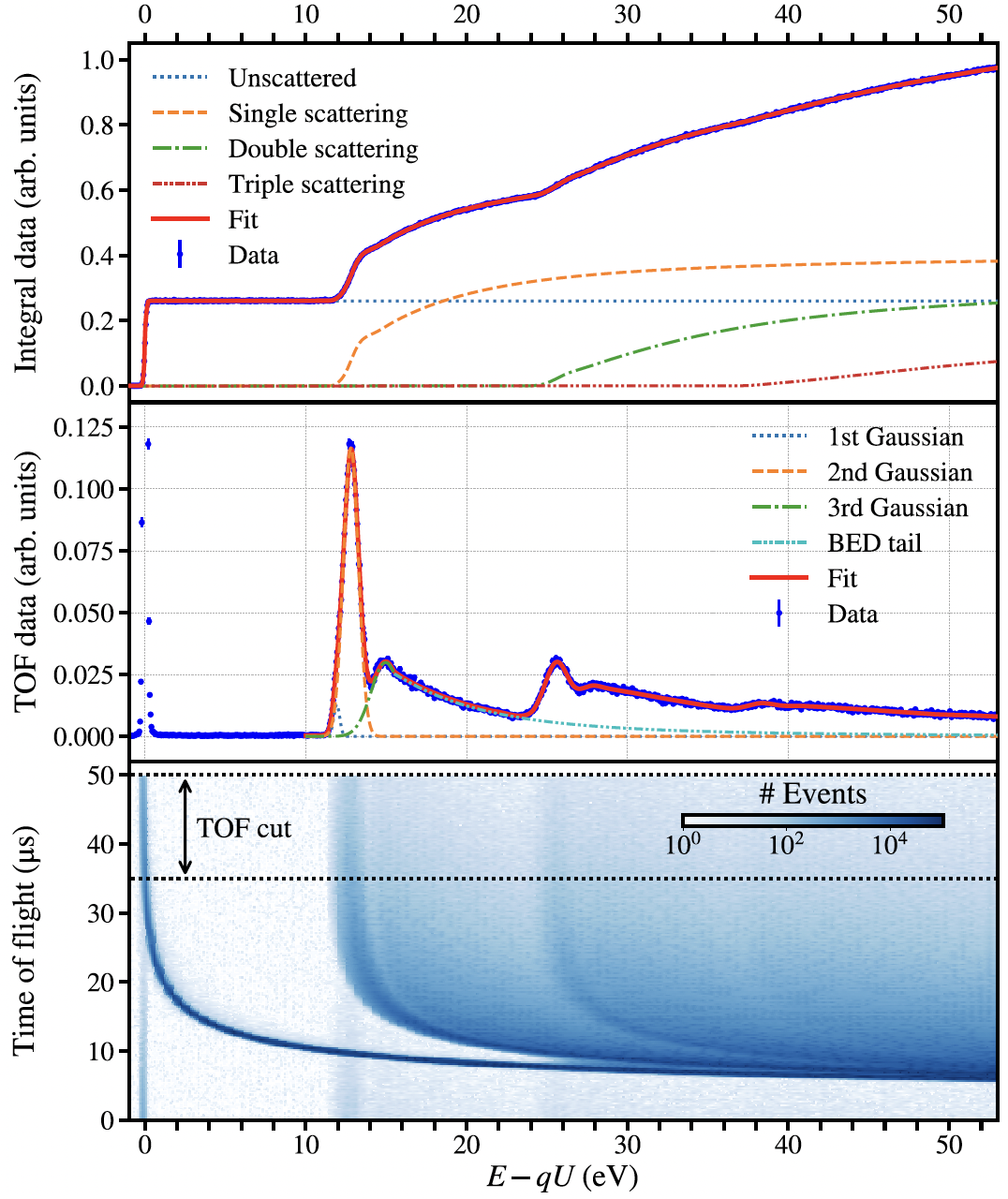}
\caption{Measurement of the response function with e-gun in the normal integral mode (top) and the time-of-flight mode (bottom) and time-of-flight selected events for $t_\mathrm{TOF}\leq 35\,\mu$s (middle) \cite{KATRIN:2021fgc} \ccbc . \label{fig:response_enloss}}
\end{figure}

This response function can be determined experimentally. To do this, electrons from the e-gun in the rear section are shot through the windowless gaseous tritium source and the count rate of these electrons transmitted through the spectrometer is measured with small steps of the retarding energy. Figure \ref{fig:response_enloss} top shows this count rate as a function of the retarding energy, here in the unit of excess energy, i.e. the start energy of the electrons minus the retarding energy. The rapid increase reflects the spectrometer's transmission function for unscattered electrons. Electrons scattered inelastically once or several times by the tritium molecules require a lower retarding energy to be counted at the detector.
The fact that the response function arises from the energy loss spectrum can also be shown by the same measurement in the differential time-of-flight mode. For this purpose, the time of flight of the electrons is measured: the electrons are created by photoionisation by a short UV laser pulse on the photocathode of the e-gun, so this laser pulse is the start signal. The stop signal for the time of flight measurement is provided by the electron detector behind the spectrometer. In Fig. \ref{fig:response_enloss} bottom, the times of flight of the electrons as a kind of hyperbola are presented as function of the excess energy. If only long times of flight above 35\,$\mu$s (Figure \ref{fig:response_enloss} middle) are accepted, only very slow electrons are selected, which can just overcome the retarding threshold, yielding a differential spectrum. In this diagram, the sharp peak of the unscattered electrons and the energy loss spectrum of the singly and multiply inelastically scattered electrons can be clearly seen. If one now integrates Fig. \ref{fig:response_enloss} centre, the response function Fig. \ref{fig:response_enloss} top is obtained also from the time-of-flight data .

\begin{figure}[t!]
\centering
\includegraphics[width=0.65\textwidth]{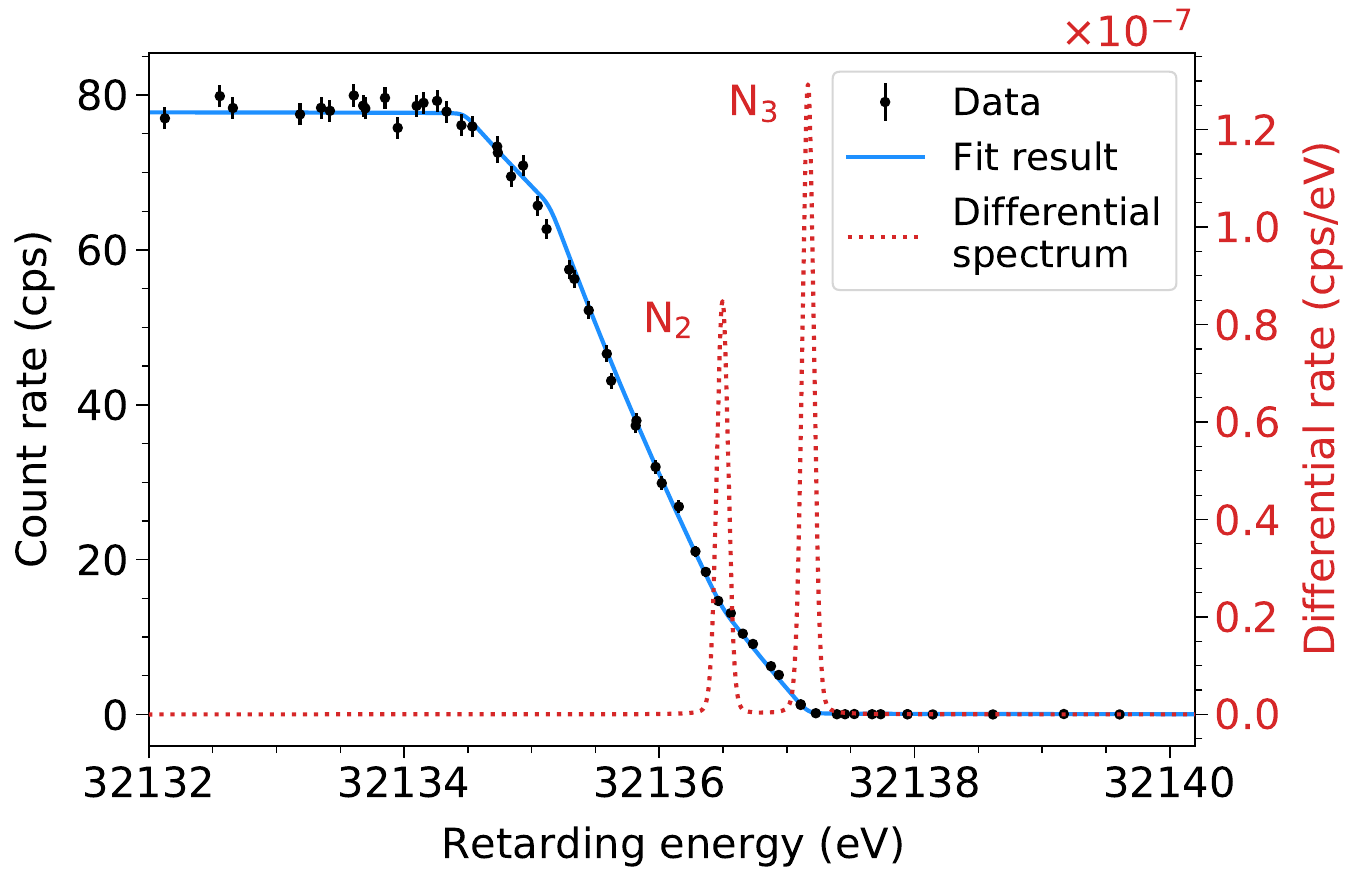}
\caption{Spectroscopy of the $\mathrm{N_{23}-32}$ conversion electron line doublet from Kr-83m cocirlating with the molecular tritium in the WGTS
\cite{KATRIN:2021uub} \ccbc .
\label{fig:KATRIN_N23-32}}
\end{figure}

For the completeness the following should be mentioned: In contrast to the \belec s from the tritium decay, the electrons from the e-gun possess a sharp angular distribution and they have to pass through the entire tritium source, while the \belec s are created 
everywhere in proportion to the tritium density in the tritium source and thus on average only traverse half of the tritium source, but some at larger angles. These changes in the response function for describing the tritium \bspec\ are applied to the response function measured by the e-gun via simulations (note differences of \emph{e-gun response} and \emph{tritium response} in Figure \ref{fig:beta_spectrum_response_convolution} bottom left). Secondly, a plasma forms in the tritium gas with the many electrons produced by \bdec\ and their subsequent inelastic scattering, as well as the associated ions. This small, Gaussian broadening of the starting potential of the \belec s cannot be detected by the response function measurement with the e-gun. Instead, it is determined from spectroscopy of the $\mathrm{N}_{23}-32$ double line of the conversion electron emitter \krm\ (see Figure \ref{fig:KATRIN_N23-32}), which is co-circulated with the molecular tritium under standard measurement conditions in the tritium source. These conversion electron lines have an extremely small line width, so that all broadening beyond the transmission function can be attributed to plasma effects. It is typically of the order ${\cal O}(30)$\,meV. A second, a slightly larger contribution to the Gaussian broadening arises from the Doppler broadening of the \belec s originating from the decay of tritium molecules at 30\,K or 80\,K.

\begin{figure}[h!]
\centering
\includegraphics[width=\textwidth]{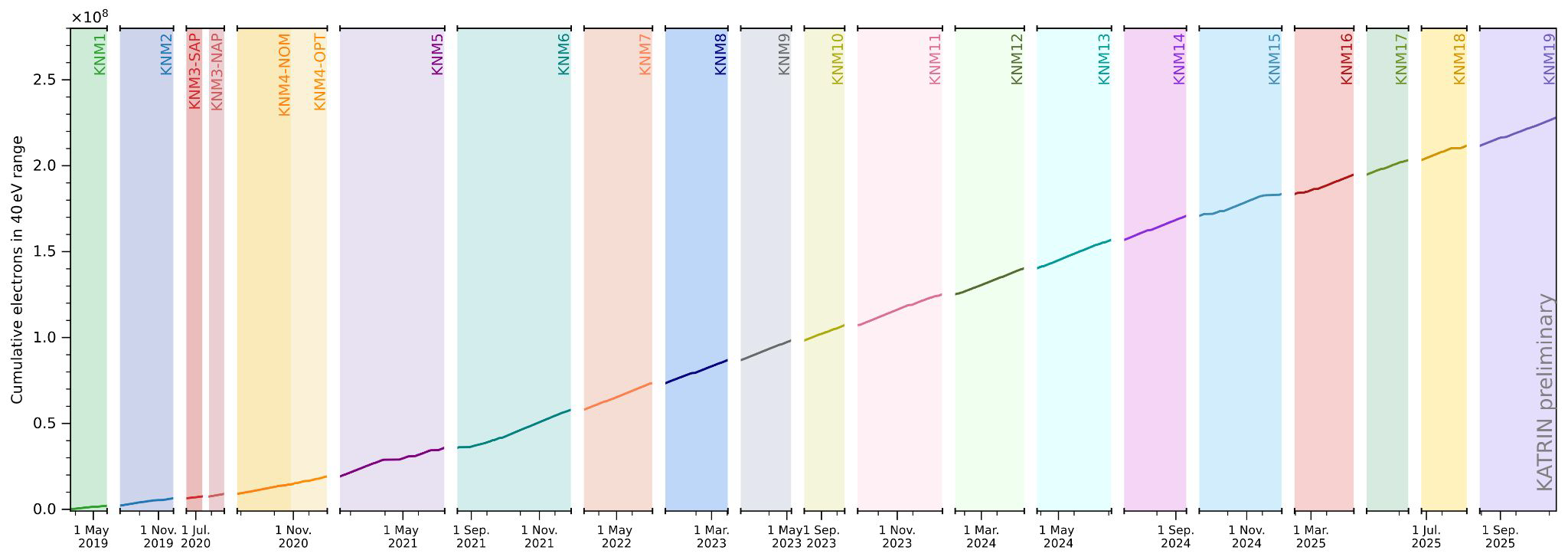}
\caption{Accumulated tritium data taking in units of counted \belec s within the last 40\,eV below the endpoint with the KATRIN experiment from 2019 to the end of 2025 with 1000 days of KATRIN data (courtesy and copyright Jaroslav \v{S}torek, KATRIN Collaboration)
\label{fig:KATRIN_runs}
}
\end{figure}

Fig. \ref{fig:KATRIN_runs} gives an overview over the various measurement runs that the KATRIN experiment has conducted since 2019 until the end of 2025 featuring the 1000-day goal of the measurement. The vertical axis shows the unique high statistics of the KATRIN experiment: In total, the spectral shape of the tritium \bspec\ in the last 40 eV below the endpoint $E_0$ will be determined with nearly 230 million electrons.

\clearpage

\section{Physics topics addressed by the experiment}\label{physics}

\begin{figure}[h!]
    \centering
    \includegraphics[width=1\textwidth]{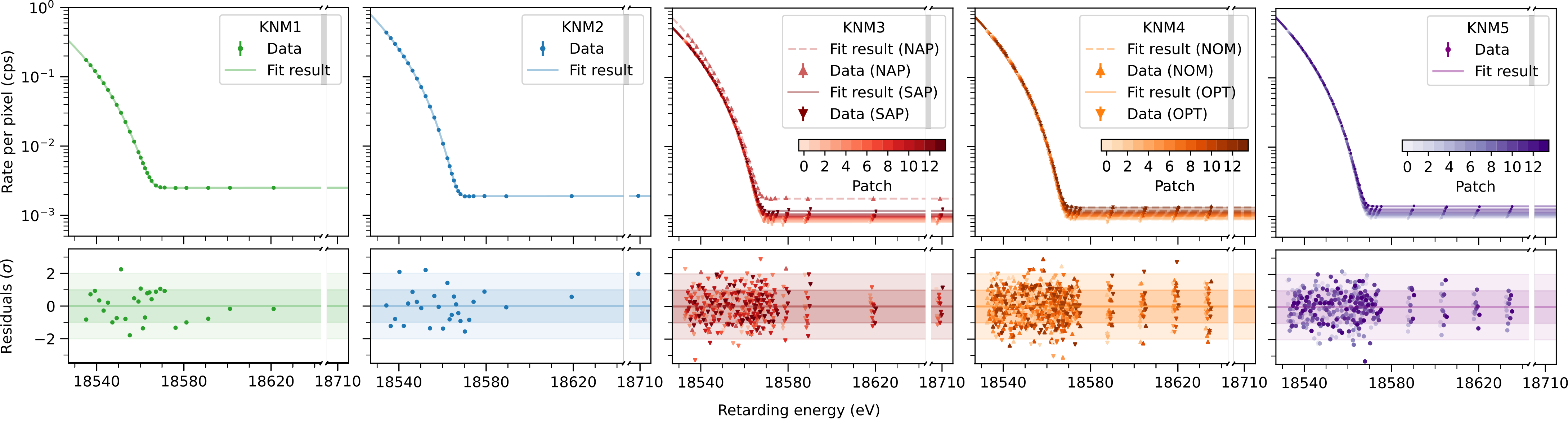}
  \caption{Experimental \bspectra\ and fits to the data of the first 5 campaigns KNM1 to KNM5 of the KATRIN experiment (upper plots) and residuals of the fits (lower plots), plot rearranged from \cite{KATRIN:2024cdt_arXiv} \ccbc , see also \cite{KATRIN:2024cdt}.
   \label{fig:beta_spectra_knm1-5}}
\end{figure}

First and foremost, the goal of the KATRIN experiment is to measure the so-called electron neutrino mass, which is an average value across all neutrino mass eigenvalues corresponding to their mixing with the electron neutrino, see Eq. \eqref{eq:electron_neutrino_mass}. To this end, a total of 1000 days of precision data in the region below the endpoint \ezero\ until the end of 2025 has been collected and is to be used to determine the neutrino mass square according to Eq. \eqref{eq:beta_spectrum2}. Fig. \ref{fig:beta_spectra_knm1-5} compares the beta spectra measured in this way for the first 5 measurement campaigns KNM1 to KNM5, which represent about one-sixth of the final statistics. From the sequence of plots it is evident that the signal strength has increased significantly since KNM2 and that, since KNM3 (SAP), the background rate has been reduced by a factor of two. The latter was achieved by shifting the analysis plane (SAP) by an amount of 5\,m towards the detector and thereby reducing the spectrometer volume downstream of the analysis plane. Although this shifted-analysis plane method leads to larger inhomogeneities in the electrical retarding potential and in the magnetic field in the curved analysis plane, this can be remedied -- 
thanks to the segmented detector and the fundamentally radially symmetric design -- by fitting 14 patch spectra, which even led to a slightly better energy resolution. Therefore, 14 measurement spectra are shown for each measurement campaign since KNM3 (SAP). Furthermore, the fitted model curves $\dot N$ with their residuals (the deviations of data points from the fit curve normalized to the statistical uncertainties) are shown, demonstrating that the model curves according to Eq. \eqref{eq:fit_fcn_beta_spec}
fit the data very well. The analysis of the measurement data was performed by minimizing the negative log-likelihood according to the frequentist method; a Bayesian analysis will be published in 2026.

\addtocounter{footnote}{-2}

The quantities $f(E,qU)$ and $b$ from Eq. \eqref{eq:fit_fcn_beta_spec} usually depend on the measurement campaigns and patches, and the various systematic uncertainties are listed in Table \ref{Tab:SystematicBreakdown} for the first measurement campaigns KNM1 to KNM5 together with uncertainties of the \bspec\ (\emph{theoretical corrections} and \emph{molecular final-states distribution}). For details, see reference \cite{KATRIN:2024cdt}.

\begin{table}[!h]
    \centering
    \caption{Breakdown of the uncertainties for the first 5 neutrino mass measurement campaigns KNM1 to KNM5. Each contribution in the last block of the table is smaller than 0.002\,eV$^2$ and they are therefore not propagated into the fit. Detailed information about the various effects is provided in the reference  \cite{KATRIN:2024cdt_arXiv} \ccbc , see also \cite{KATRIN:2024cdt}.}
    \begin{tabular}{lc}
        \hline   \textbf{Effect} & \textbf{68.3\,\%} CL uncertainty on $\boldsymbol{m_\nu^2}$ (eV$^2$) \\ \hline 
        Statistical uncertainty                               & $0.108$ \\ 
        Non-Poissonian background                             & $0.015$ \\ \hline 
        Column density $\times$ inelastic cross section       & $0.052$ \\ 
        Energy-loss function                                  & $0.034$ \\ 
        Scan-step-duration-dependent background               & $0.027$ \\ 
        Source-potential variations                           & $0.022$ \\
        $qU$-dependent background slope                       & $0.007$ \\ 
        Analyzing-plane magnetic field and potential          & $0.006$ \\ 
        Source magnetic field                                 & $0.004$ \\ 
        Maximum magnetic field                                & $0.004$ \\
        Rear-wall residual tritium background                 & $0.004$ \\ \hline 
        Molecular final-state distribution                    & $< 0.002$ \\
        Activity fluctuations                                 & $<0.002$\\
        Detector efficiency                                   & $<0.002$\\
        Retarding-potential stability and reproducibility   &  $<0.002$ \\
        Theoretical corrections                               & $<0.002$\\  \hline 
    \end{tabular}
    \label{Tab:SystematicBreakdown}
\end{table}

A data blinding scheme to avoid bias was implemented in the analysis of the KATRIN data.
For this purpose, the molecular final-states distribution $(V_j, W_j)$, see Eq. \eqref{eq:beta_spectrum2}, is convolved with a Gaussian variance that is kept hidden until the final unblinding. Furthermore, all analyses in the first phase are performed using Monte Carlo twins, which incorporate the real experimental parameters and the values of the countless sensors of the KATRIN experiment. The data analysis is performed independently by two different analysis teams using different fit programs. The influence of systematic errors is also analyzed using different methods (covariance matrix, pull terms, Monte Carlo method) in order to achieve robustness and reliability in the results.
The main four fit parameters are the normalisation parameter $A_\mathrm{s}$, the endpoint energy \ezero , the background rate $b$, and the neutrino mass square \mnuetwo . 

\addtocounter{footnote}{-2}

It should be noted that the KATRIN Collaboration fits the parameter \mnuetwo\ without constrains, i.e. also negative values of \mnuetwo\ are allowed although these are unphysical. This is done to take  statistical fluctuations of the data into account, since a positive value of \mnuetwo\ shifts the spectra upwards as seen in the insert of Fig.  \ref{fig:tritium_beta_spectrum_at_endpoint} right, whereas a negative value of \mnuetwo\ shifts it downwards. 

\begin{figure}[t]
    \centering
    \includegraphics[width=0.6\textwidth]{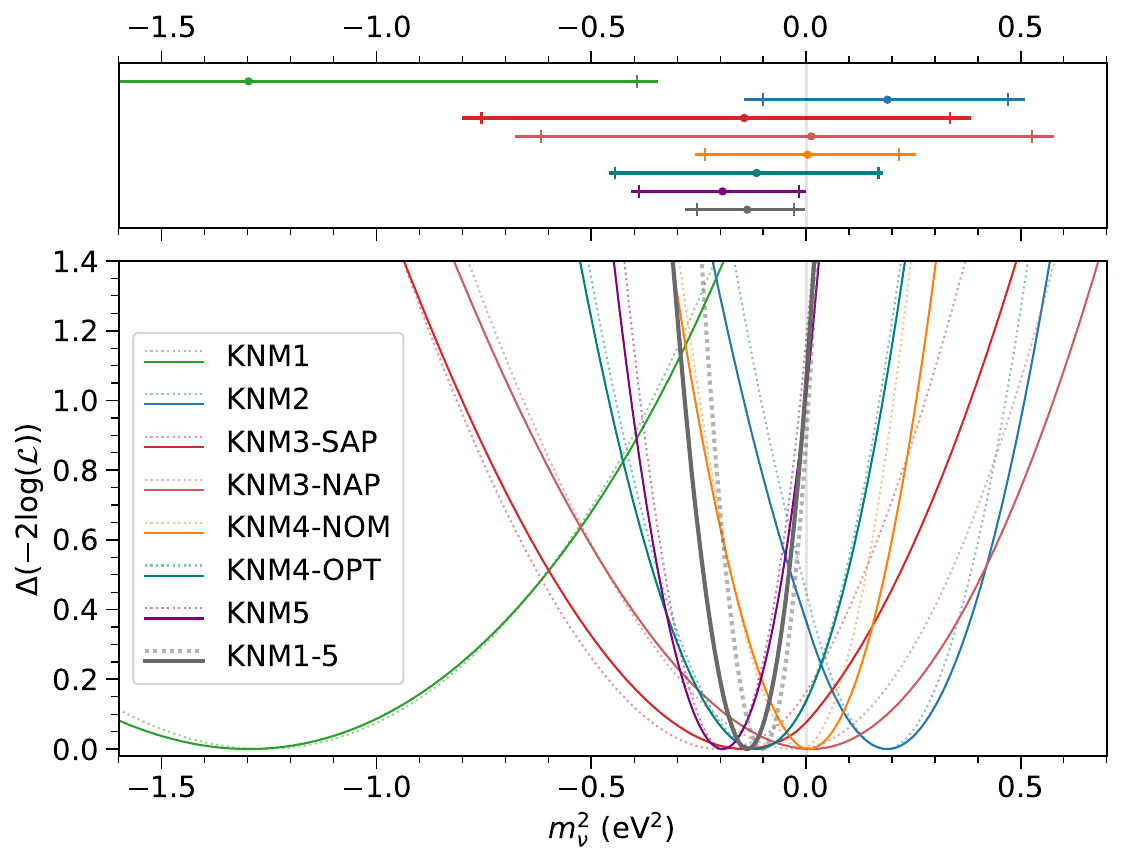}
    \label{fig:log_likelihood_knm1-5}
    \caption{Negative log-likelihood curves for the different campaigns KNM1 to KNM5. Please note that campaigns KNM1 and KNM2 were reevaluated with the newest inputs and differ slightly from references \cite{KATRIN:2019yun,KATRIN:2021uub}. The campaign KNM3 was split into the part with the shifted analysis plance (SAP) and with the symmetric analysis plane (NAP), and the campaign KNM5 had to be split because of an energy scale drift into two phases NOM and OPT \cite{KATRIN:2024cdt_arXiv} \ccbc , see also \cite{KATRIN:2024cdt}.}
\end{figure}

Fig. \ref{fig:log_likelihood_knm1-5} shows the different negative log-likelihood curves 
from the fitted unblinded experimental data
around the respective minimum as a function of the neutrino mass square \mnuetwo .
The various data sets are compatible with each other and deliver a common value for \mnuetwo\ of
\begin{equation}
  m^2_{\nu, \mathrm{KNM1-5}} = -0.14^{+0.13}_{-0.15}~\mathrm{eV^2} \label{eq:knm1-5_result}
\end{equation}
This value includes statistical and systematic uncertainties and is compatible with a vanishing neutrino mass from which an upper limit on the neutrino mass scale $m_\nu$ can be deduced.

\addtocounter{footnote}{-2}

\begin{figure}[h!]
    \centering
    \includegraphics[width=0.5\textwidth]{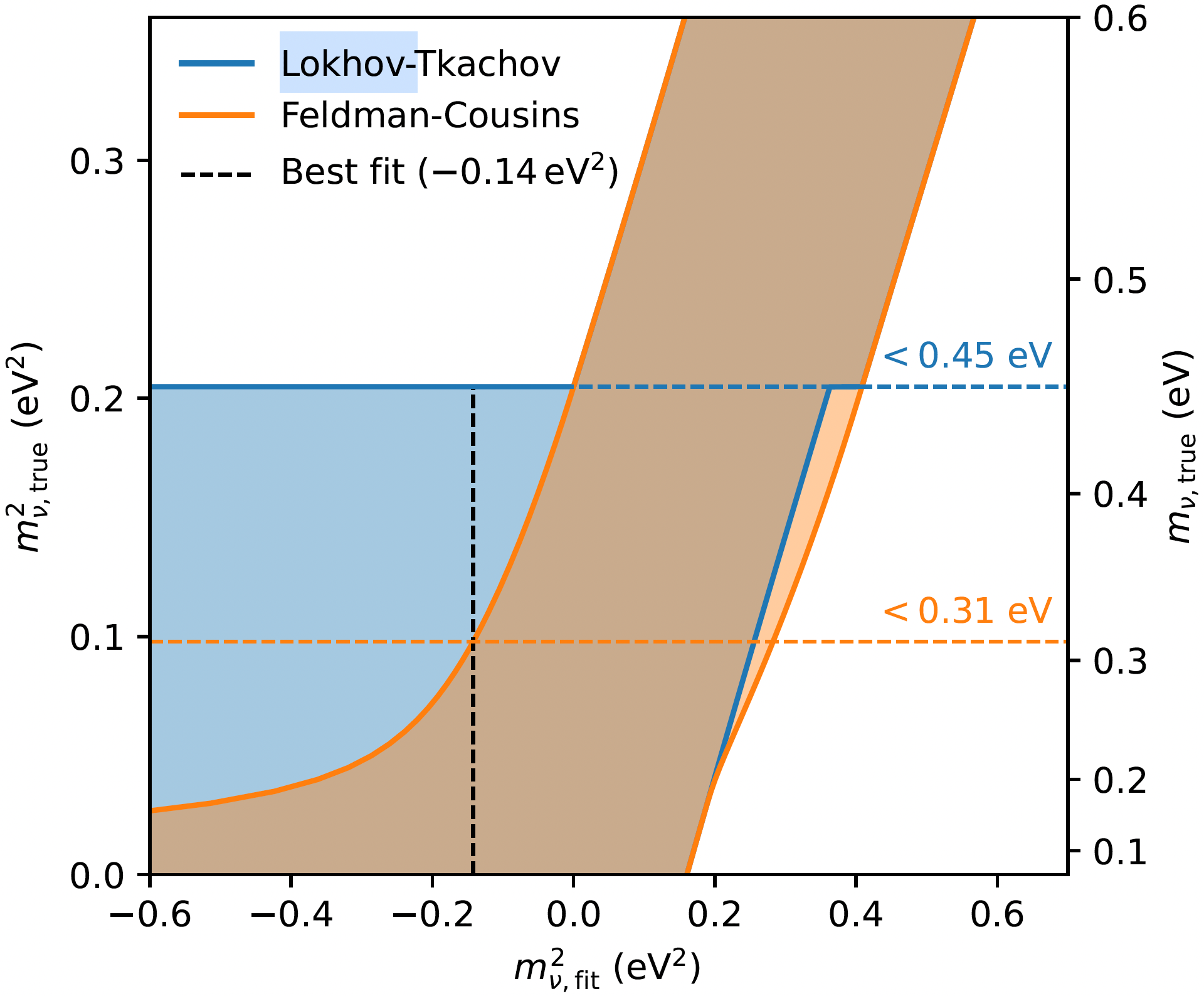}
    \caption{Upper limit setting for the result from Eq. \eqref{eq:knm1-5_result} by the Lokhov and Tkachov \cite{Lokhov:2014zna} (blue) and by the Feldman and Cousins \cite{Feldman:1997qc} (yellow) methods. Both methods overlap for large positive values of $m^2_\mathrm{\nu,~fit}$ but differ for negative values ones \cite{KATRIN:2024cdt_arXiv} \ccbc , see also \cite{KATRIN:2024cdt}.  \label{fig:upper_limit}}
\end{figure}

Fitting the parameter \mnuetwo\ without constraints has to account for the fact that any unaccounted broadening $\sigma^2$ that is incorporated into the measurement data
will bias the \mnuetwo\ value obtained in the fit via the relation $\Delta m^2_\nu = - 2 \sigma^2$, thereby pushing the result in the direction of negative values \cite{Robertson:1988xca}.
The latter effect and the fit of the parameter \mnuetwo\ without restriction motivated the KATRIN collaboration to exercise particular caution when calculating the neutrino mass limit.
While the standard method in the field, developed by Feldman and Cousins \cite{Feldman:1997qc}, leads to more stringent limits the more negative the \mnuetwo\ value becomes, the method developed by Lokhov and Tkachov \cite{Lokhov:2014zna}
always gives the same limit for negative \mnuetwo\ values, namely the sensitivity. Due to the above-mentioned risk of unintentional bias due to unknown additional variance, the KATRIN Collaboration decided to adopt the method by Lokhov and Tkachov as shown in Fig. \ref{fig:upper_limit}:
\begin{equation}
  m_{\nu, \mathrm{KNM1-5}} < 0.45 ~\mathrm{eV~ (90\,\% ~C.L.)} 
\end{equation}
The upper limit of Feldman and Cousins would be $m_{\nu, \mathrm{KNM1-5}} < 0.31 ~\mathrm{eV~ (90\,\% ~C.L.)}$.
At the end of the measuring phase with about 6 times more data recorded, the KATRIN final sensitivity on the neutrino mass will be $< 0.30$\,eV at (90\,\% ~C.L.).

The unprecedented precision data of the \bspec\ below the endpoint from KATRIN can also be used to perform other searches for physics beyond the Standard Model of particle physics.
\begin{itemize}
\item {\bf Search for eV sterile neutrinos:} Previous persistent anomalies in neutrino oscillation experiments with accelerator and reactor neutrinos, as well as in the calibration of solar neutrino experiments with gallium, suggested that there might be a fourth neutrino with a mass of ${\cal O}(1)$\,eV. However, as measurements of the width of the weak gauge boson Z$^0$ show, this fourth neutrino cannot couple to the weak gauge bosons W$^\pm$ and Z$^0$, so it must be sterile. Nevertheless, it can become physically manifest in neutrino oscillations or in \bdec\ if it mixes sufficiently strong with the other active neutrinos, i.e., so that the neutrino mixing matrix $U_{\alpha i}$ can be extended to a $4 \times 4$ matrix. 

To investigate this question, several short baseline reactor neutrino oscillation experiments were carried out worldwide. One of which, Neutrino-4, claimed to have found this fourth neutrino. These short baseline reactor neutrino experiments are particularly sensitive to neutrino mass squares in the range of $\Delta m^2_{41} \leq 10\,\mathrm{eV^2}$, i.e., to mass states of the fourth neutrino of $m_4 \leq 3$ eV. By contrast, the KATRIN experiment has a complementary sensitivity to neutrino masses $m_4 \geq 0.5$ eV, because the signature of a sterile neutrino mass state would be the appearance of another branch in the \bspec\ that would lead to a kink at $E_0-m_4$ below the endpoint (see Fig. \ref{fig:beta_spec_sterile_eV}).
The \bdec\ branch into the three light neutrino mass states is described by  Eq. \eqref{eq:electron_neutrino_mass} with a multiplicative normalization of $\cos^2{\Theta_\mathrm{ee}}$, whereas the \bdec\ branch into the fourth neutrino mass state is described by the fraction 
$\sin^2{\Theta_\mathrm{ee}}$:
\begin{eqnarray} \nonumber
\dot N(E) \propto \hspace*{1cm} &~ \hspace*{-2.6cm} F(A,Z+1) \cdot p_\mathrm{e} \cdot (E+m) \cdot \sum_j W_j \cdot (E_0-E-V_j) \\ \nonumber
\cdot \hspace*{0.15cm} \biggl(\hspace*{-0.2cm} &~ \cos^2{\Theta_\mathrm{ee}}
\cdot \sqrt{(E_0-E-V_j)^2 - m^2_\nu} \cdot \Theta(E_0-E-V_j-m_\nu)\\ 
&+
\sin^2{\Theta_\mathrm{ee}}
\cdot \sqrt{(E_0-E-V_j)^2 - m^2_4} \cdot \Theta(E_0-E-V_j-m_4) \biggr)
\label{eq:beta_spectrum_eVsterile}
\end{eqnarray}

\begin{figure}[th]
    \centering
    \includegraphics[width=0.5\textwidth]{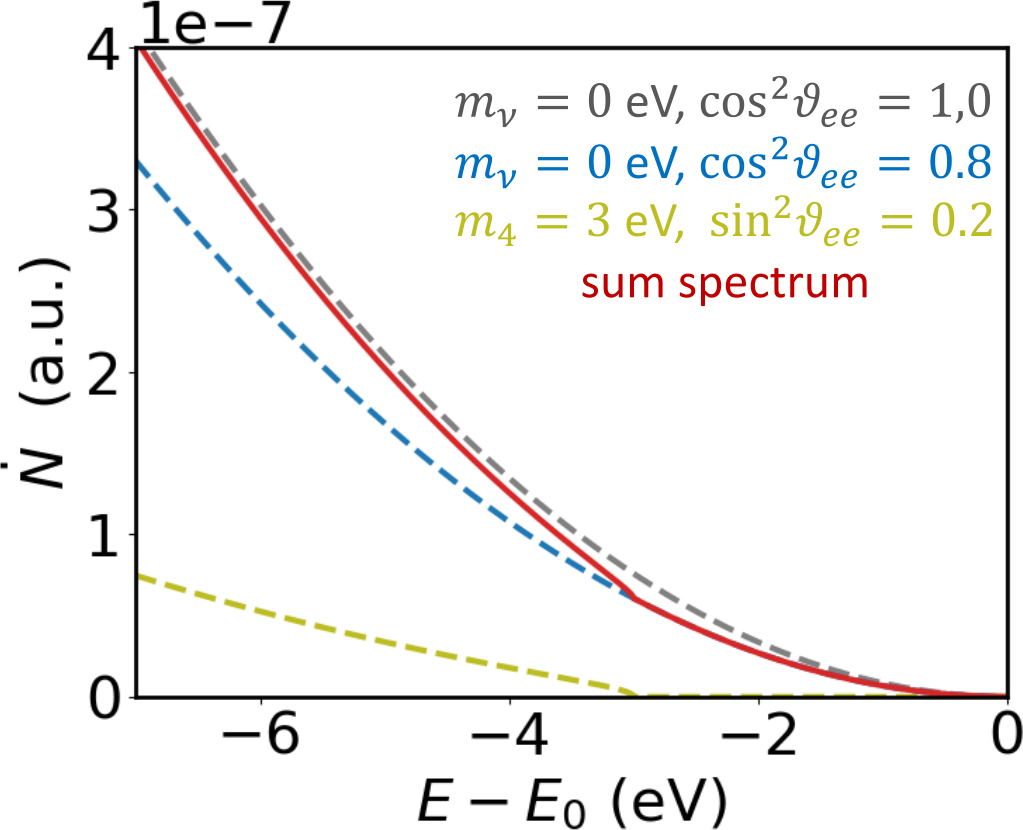}
    \caption{Beta spectrum of tritium near its endpoint for a vanishing light neutrino mass of $m_\nu=0$ (dashed gray and blue) and a rather large ($\sin^2{\Theta_\mathrm{ee}}=0.2$) admixture of a sterile neutrino with $m_4=3$\,eV (dashed ocher) and the sum spectrum of the two \bdec\ branches (red), copyright Christian Weinheimer.  \label{fig:beta_spec_sterile_eV}}
\end{figure} 

The KATRIN data of campaigns KNM1 to KNM5 were scanned for the signal of a fourth sterile state using Eq. \eqref{eq:beta_spectrum_eVsterile} with the assumption of $m_\nu=0$ \cite{Acharya2025}. 
As no additional kink was found this allows to exclude a large parameter range of neutrino masses and neutrino mixing angles complementary to the short baseline reactor neutrino experiments, It clearly refuted the claim of the Neutrino-4 experiment, see Fig. \ref{fig:sterile_eV_neutrino_exclusion}.

\begin{figure}[th]
    \centering
    \includegraphics[width=0.8\textwidth]{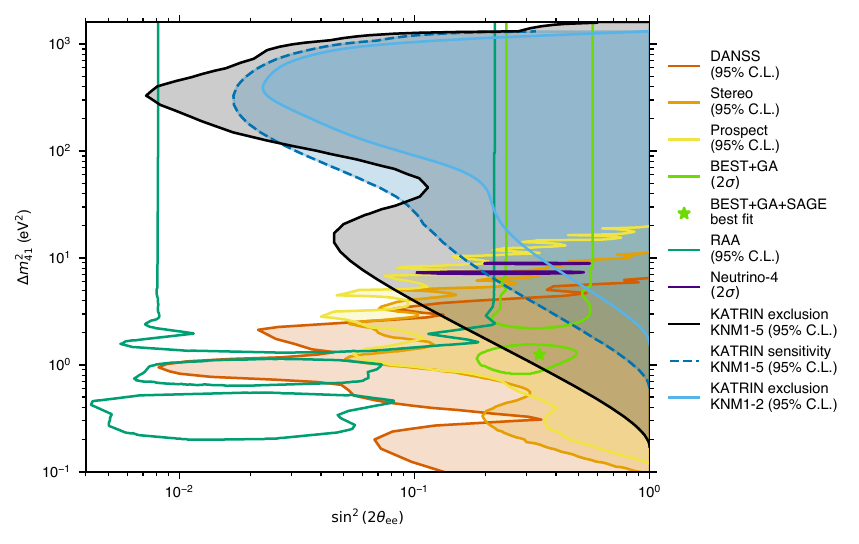}
    \caption{Exclusion limits at 95\% C.L. by the KATRIN KNM1-5 data (black) for sterile neutrinos in the mixing angle and squared mass difference plane, complementary to exclusions by short baseline reactor neutrino experiments DANSS (red), Stereo (orange) and Prospect (yellow). The open contours illustrate possible signals for the reactor antineutrino anomaly (dark green), the BEST experiment together with the gallium anomaly and the solar neutrino experiment SAGE (light green) and the short paseline reactor neutrino experiment Neutrino-4 (magenta), \cite{Acharya2025} \ccbc , please see references therein.      \label{fig:sterile_eV_neutrino_exclusion}}
\end{figure} 

\item{\bf Search for sterile keV neutrinos:}
Fig. \ref{fig:sterile_eV_neutrino_exclusion} shows that the standard KATRIN data are sensitive to squared neutrino mass differences $\Delta m^2_{41}<2000$ eV$^2$, 
i.e., to sterile neutrino masses of $m_4 \leq 40$ eV. At the very beginning of the KATRIN measurement campaigns, the KATRIN collaboration conducted an exploratory study in which it measured the \bspec\ significantly deeper below the endpoint and searched for sterile keV neutrinos up to masses of 1\,keV \cite{KATRIN:2022spi}. 
In the range $0.1\,\mathrm{keV} \leq m_4 \leq 1\,\mathrm{keV}$, previous upper limits on possible neutrino mixing could be improved by about an order of magnitude down to a few 
$10^{-4}$. Since sterile neutrinos of a few keV have been proposed as warm dark matter candidates, this mass range will be specifically investigated in a dedicated phase 2 of the KATRIN experiment (TRISTAN phase) following an upgrade of its current read-out system to the TRISTAN detector (see section \ref{future}).

\item {\bf Search for the cosmic neutrino background:}
After neutrinos decoupled approximately 1 second after the Big Bang, the matter-energy budget of the universe comprises also a cosmic neutrino background (C$\nu$B) with a density of 336 neutrinos per cubic centimeter, summed over all neutrino types \cite{Lesgourgues:2013sjj}, similar to the cosmic microwave background of photons (CMB). The C$\nu$B is indirectly visible as relativistic degrees of freedom in both the element distribution from the Big Bang nucleosynthesis and the power spectrum of the cosmic microwave background. A $\nu_\mathrm{e}$ from the C$\nu$B can be captured by a tritium nucleus emitting a monoenergetic electron in inverse beta decay:
\begin {equation}
\nu_e + \mathrm{T} \rightarrow \mathrm{^3He^+} + \mathrm{e^-}
\end{equation}
The monoenergetic electron features a kinetic energy of $E_0 + m_\nu$, so it appears separated from the actual \bspec\ by the small amount of $2\cdot m_\nu$. However, the separation disappears for neutrino masses approaching zero. Additionally, the monoenergetic electron line is smeared by the energy resolution of the experiment and, in our case of tritium molecules, by the rotational broadening of the final state (see \eqref{eq:beta_spectrum2}).  In order to obtain a measurable signal, the KATRIN tritium source would need to be upscaled by many orders of magnitude of tritium molecules, but the inverse \bdec\ signal 
would probably still be buried under the background rate $b$. Looking for a C$\nu$B signal, the KATRIN experiment has so far even then been able to set an upper limit on the local relic neutrino overdensity ratio of $\eta < 9.7 \cdot 10^{10}/\alpha$ at a confidence level of 90\% with $\alpha = 1 (0.5)$ for Majorana (Dirac) neutrinos
\cite{KATRIN:2022kkv}.

\item {\bf Search for Lorentz invariance violation:}
The $CPT$ theorem states that a Lorentz-invariant, local, unitary quantum field theory, such as the Standard Model to describe electroweak and strong interactions, must be invariant under the combined transformation of charge conjugation ($C$), parity inversion ($P$), and time reversal ($T$). It is therefore interesting to look for possible small violations of Lorentz invariance \cite{PhysRevD.69.105009}. This could manifest itself, for example, in a sidereal oscillation of the shift of the endpoint \ezero\ of a \bdec\ \cite{LEHNERT2022137017}.
The KATRIN Collaboration has searched for such a signature, the non-observation allows the experiment to
set new limits on certain parameters of Standard Model extensions \cite{KATRIN:2022qou}.

\item {\bf Investigation of general neutrino interactions:}
The shape of the tritium \bspec\ of Eq. \eqref{eq:beta_spectrum_simple} follows from the low energy \emph{V-A} effective theory of the Standard Model mainly governed by the phase space of the out-going neutrino and electron. Additional  scalar, pseudo and tensor terms in the Lagrangian of \bdec s have been proposed \cite{PhysRev.106.517} and investigated since long \cite{Severijns:2011zz}, e.g. by searching for deviations the \bspec s shape. With the above mentioned, well movitated possibility of additional sterile neutrinos even more terms in the Lagragian could become possible \cite{Ludl:2016ane}. The KATRIN data of the second measurment campaign KNM2 have been investigated for the occurance of shape variations by additional terms in the Lagrangian and new limits on some general neutrino interaction parameters have been set \cite{KATRIN:2024odq}.

\item {\bf Other physics results from KATRIN:} 
In addition to the aforementioned searches for the active neutrino mass, sterile neutrino states at the eV- and keV-scale and other phenomena beyond the Standard Model, the KATRIN experiment, with its highest requirements for measurement accuracy and stability as well as novel technologies, contributes to many new insights in physics and technical innovations. Examples include precision spectroscopy of \krm\ conversion electrons, laser Raman spectroscopy of hydrogen isotopologues, investigations of thin plasmas, precise tracking of charged particles in electromagnetic fields, final states of \bdec s of tritium molecules, ppm-precise high voltage, extreme ultra-high vacuum, \dots.
\end{itemize}

\clearpage

\section{Complementarity with other experiments in the field 
}
\label{compl}
\subsection{Other direct neutrino mass experiments}


Other experiments and projects are being carried out that directly search for the neutrino mass from the kinematics of a weak decay such as tritium \bdec\ or electron capture at $^{163}$Ho. More information can be found in the following review articles \cite{Mertens:2014nha,Nucciotti:2015rsl,Formaggio:2021nfz}. These experiments and projects are in the process to develop very interesting technologies, but, in general,  have not yet reached the sensitivity of the KATRIN experiment:
\begin{itemize}
\item Similar to KATRIN, the \emph{Project 8} experiment uses a molecular gaseous tritium source in a high magnetic field \cite{Project8:2023jkj}. However, the \bspec\ is measured in a completely different way: part of the emitted \belec s is captured by magnetic mirrors  according to Eq. \eqref{eq:adiabatic_invariant} by increasing the magnetic field strength at the end of the tritium source.  The electrons gyrating in the magnetic field emit cyclotron radiation which is collected by a waveguide, mixed down and very sensitively detected with the help of a  cryogenic amplifier.
The frequency of the cyclotron radiation is inversely proportional to the relativistic $\gamma$ factor of the \belec , so that the energy $E$ of a \belec\ can be determined by measuring its cyclotron frequency. Using this method, known as \emph{cyclotron radiation emission spectroscopy} (CRES) \cite{Monreal:2009za}, Project 8 measured a first tritium \bspec\  and an upper neutrino mass limit of $m_\nu \leq 155$\,eV was achieved \cite{Project8:2022hun}.
 The main advantage of measuring cyclotron emission instead of \belec s directly is that a tritium source with an even higher column density than in the KATRIN experiment becomes opaque to \belec s -- due to the inelastic scattering of \belec s on tritium molecules -- but not to cyclotron radiation.
 In further planned project stages \cite{Project8:2022wqh}, the statistics shall be significantly improved by a much larger setup and a  resonant cavity for detecting cyclotron radiation.
By using atomic tritium, the systematic uncertainties shall be kept so small that a final neutrino mass sensitivity of 40 meV can be achieved.
\item The project \emph{Quantum Technologies for Neutrino Mass} (QTNM) also aims to use CRES technology with a slightly different concept \cite{Amad:2024jod}.
\item  
The \emph{PTOLEMY} project \cite{PTOLEMY:2018jst} was originally launched to measure cosmic background neutrinos via the inverse \bdec\ but now it also aims to perform high-precision spectroscopy of the tritium \bspec\ at the end point for direct determination of the neutrino mass. It uses tritiated graphene as a tritium source. High-resolution transition edge sensors (TES) are to be used to determine the energy of the \belec s, but these can only achieve their highest energy resolution if the electrons are specifically decelerated beforehand. This is to be done in a novel transverse electromagnetic filter \cite{Betti:2018bjv}, which is targeted to decelerate each electron specifically after its transverse energy has been determined using CRES technology. The experiment has begun to demonstrate the first of its new ideas experimentally \cite{PTOLEMY:2025unk}. 
\item Not only is the endpoint range of a \bspec\ sensitive to the neutrino mass, but also electron capture. At first, only an electron neutrino is emitted, but the captured electron leaves a hole in the electron shell, which is refilled resulting in X-ray transitions or Auger electrons. This electromagnetic de-excitation spectrum consists mainly of Lorentz-shaped lines with wide tails, which are limited to high energies by the maximum available energy. There, the de-excitation spectrum is essentially limited by the phase space of the emitted neutrino
$E_\mathrm{tot,\nu} \cdot p_\nu = E_\mathrm{tot,\nu} \cdot \sqrt{E^2_\mathrm{tot,\nu} - m^2_\nu}$
and thus sensitive to the neutrino mass square \mnuetwo , just as in the case of the endpoint region of a \bspec . Two experiments are dedicated to this possibility with the electron-capture isotope $^{163}$Ho. Since not only one particle but all electromagnetic de-excitations must be measured, these experiments use arrays of cryogenic bolometers in which the $^{163}$Ho is implanted. The \emph{ECHo} experiment \cite{Gastaldo:2017edk} uses metallic microcalorimeters (MMC) as bolometers, which are read out via SQUIDs.
First $^{163}$Ho de-excitation spectra were measured within the first phase of ECHo, ECHo-1k, using approximately 50 individually read-out MMCs with an energy resolution of 6.6 eV (FWHM). A neutrino mass limit of less than 15 eV (90\,\% C.L.) was reported \cite{lqkb-hylx}. 
In the next expansion stage of ECHo, it is expected that orders of magnitude more MMCs are to be used with application of frequency multiplexing \cite{Richter:2021urk}, which allows significantly more MMCs to be read out with a single SQUID amplifier.
\item 
The \emph{HOLMES} experiment is similar to ECHo. It also investigates electron capture of $^{163}$Ho implanted in an array of cryogenic calorimeters. However, HOLMES uses an array of transition-edge sensor (TES) microcalorimeters, achieving an average energy resolution of 6 eV FWHM with a scalable, multiplexed readout technique. With the first de-excitation spectrum after electron capture at $^{163}$Ho, HOLMES achieved an upper limit on the neutrino mass of 27\,eV (90\,\% C.L.) \cite{Alpert:2025tqq}. Strictly speaking,  results from ECHo and HOLMES and those obtained by KATRIN and Project 8 measure different quantities. While the latter determine the effective mass of the electron antineutrino, electron capture is sensitive to the effective mass of the electron neutrino, which, however, should yield identical values under CPT conservation.
\end{itemize}

\subsection{Neutrinoless double beta decay}
\begin{figure}[th]
    \centering
    \includegraphics[width=0.85\textwidth]{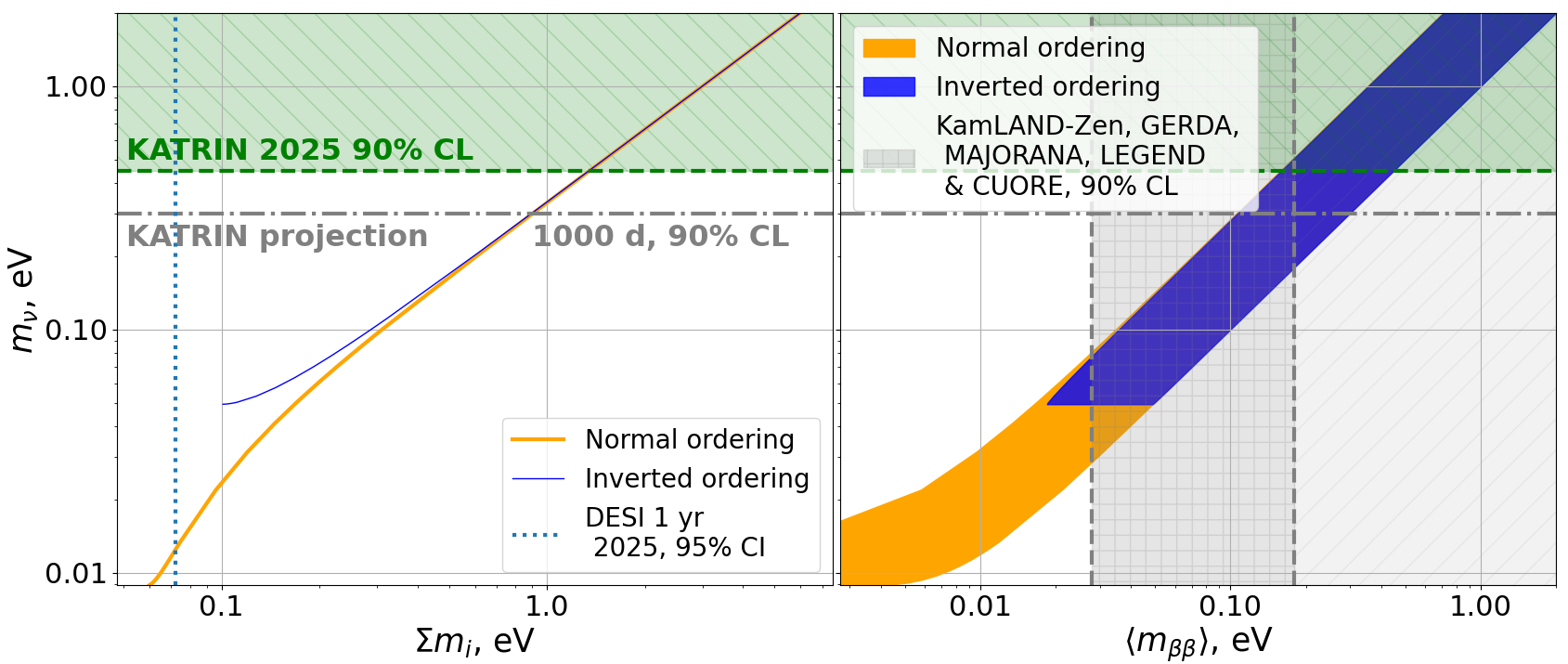}
    \caption{Connection of possible values of the effective neutrino mass from \bdec\ $m_\nu$ (Eq. \eqref{eq:electron_neutrino_mass}) as function of the sum of all neutrino masses $S_\nu$  (Eq. \eqref{eq:sum_neutrino_masses}, left) and of the effective neutrino mass from double beta decay $m_{\beta\beta}$ (Eq. \eqref{eq:effective_mass_0nbb}, right) for normal (orange) and inverted neutrino mass ordering (blue) using the constraints on quadratic neutrino mass differences 
    and on neutrino mixing angles derived in neutrino oscillation experiments.
 (adapted from \cite{Lokhov:2022zfn}, courtesy and copyright Alexey Lokhov)\label{fig:neutrino_mass_different_methods}}
\end{figure}

Specific atomic nuclei with an even number of both protons and neutrons cannot energetically undergo a simple \bdec\ or electron capture, but can undergo a double beta decay, in which -- in the case of  $\beta^- \beta^-$ decay -- two neutrons in the same atomic nucleus simultaneously transform into two protons, emitting each an electron and an electron antineutrino. Such a second order weak decay is characterized by typical half-lives in the range of $10^{20}$ to $10^{22}$ years. In case that the neutrinos are identical to their antiparticles (\emph{Majorana neutrinos}), which is not prohibited by charge conservation due to their neutrality, neutrinoless double beta decay is possible: Here the neutrino state emitted at one vertex is absorbed at the other vertex. Thus, the two electrons carry away the entire decay energy neglecting the recoil on the nucleus, please see reviews \cite{Henning:2016fad,Dolinski:2019nrj,Agostini:2022zub,Gomez-Cadenas:2023vca}. Due to the left-handedness of weak charged currents, this hypothetical process is sensitive to the neutrino mass, more precisely to the following coherent sum of neutrino mass eigenstates weighted by the neutrino mixing matrix:
\begin{equation}
m_{\beta\beta}= |\sum_i U^2_{ei} \cdot m_i |
\label{eq:effective_mass_0nbb}
\end{equation}
Here, the squares of the matrix elements $U^2_{ei}$ are used, but in contrast to the effective electron neutrino mass from \bdec\ Eq. \eqref{eq:electron_neutrino_mass}, their absolute values are not taken into account. Therefore, cancellations may occur in the effective neutrino mass from the neutrinoless double beta decay $m_{\beta\beta}$, because the matrix elements of the unitary neutrino mixing matrix are generally complex-valued. Fig. \ref{fig:neutrino_mass_different_methods} shows the connection of possible values of the effective neutrino mass from \bdec\ $m_\nu$ as function of the sum of all neutrino masses $\sum_i m_i$ (see Sec. \ref{cosmology}) (left) and of the effective neutrino mass from double beta decay $m_{\beta\beta}$ (right)  using the constraints on squared mass differences and on mixing angles derived in neutrino oscillation experiments \cite{Lokhov:2022zfn}.
Various experiments employing different isotopes actively search for neutrinoless double beta decay, but it has not yet been found. Currently, the lower limits of the half-life for neutrinoless double beta decay are a few $10^{26}$ years. For the limits on the effective neutrino mass from double beta decay, a range is usually given, as there is not only the uncertainty of the mixing angles and complex phases of the neutrino mixing matrix elements $U_{\mathrm{e}i}$ but also there exist significant uncertainties of the nuclear matrix elements $M_{0\nu}$ \cite{Engel:2016xgb,Fang:2018tui} which enter the relation \eqref{eq:dbd} of half-life $T^{0\nu}_{1/2}$ and effective neutrino mass with $m$ being the electron mass and $G_{0\nu}$ the phase space factor:
\begin{equation}\label{eq:dbd}
\frac{1}{T^{0\nu}_{1/2}} = \frac{m^2_{\beta\beta}}{m^2} \cdot G_{0\nu} \cdot |M_{0\nu}|^2
\end{equation}

\subsection{Neutrino mass from cosmology}
\label{cosmology}

Because the number density of the C$\nu$B, at 336 per cm$^3$, is similar to the density of photons in the \emph{cosmic microwave background} (CMB) but approximately 1 billion times larger than that of atoms in the universe, neutrinos do play a role in the formation of the structure of the universe despite their very small mass. The influence of non-zero neutrino masses is imprinted in both the power spectrum of the CMB and in the \emph{baryon acoustic oscillations} (BAO) of large scale structures \cite{Lesgourgues:2013sjj}. Typically, precise data from both CMB and BAO observables, as available today from the satellite missions PLANCK and DESI or EUCLID, are used to make conclusions about neutrino masses, more precisely the gravitationally effective sum of all neutrino masses 
\begin{equation}
    \label{eq:sum_neutrino_masses}
       S_\nu := \sum_i m_i
\end{equation}
The recent DESI survey of the distribution of galaxies in the universe, combined with Planck's CMB data and the standard model of cosmology, the cold dark matter model with cosmological constant ($\Lambda$CDM), yields a rather stringent upper limit on neutrino mass 
\cite{w9pk-xsk7} of 
\begin{equation}
S_\nu < 0.064\,\mathrm{eV} ~\mathrm{(95\,\% C.L.)}
\end{equation}
However, the likelihood exhibits its maximum in the unphysical, negative neutrino mass range, so that the model has weaknesses in describing the data. 
Also, there is an ongoing discussion as to whether dark energy is merely the cosmological constant $\Lambda$ introduced by A. Einstein, or whether its generic parameter $w$, which defines the ratio of pressure to energy density, does assume a value other than -1, or whether dark energy may even be time-dependent.
Such considerations beyond $\Lambda$CDM are supported by the additional problem that the determinations of the Hubble constant using data from the early and late universe do differ significantly, so that cosmologists are now calling for the neutrino mass scale to be measured in the laboratory in order to provide input for cosmological analyses.

\clearpage

\section{Future development} 
\label{future}

\begin{figure}[h]
    \centering
    \includegraphics[width=0.8\textwidth]{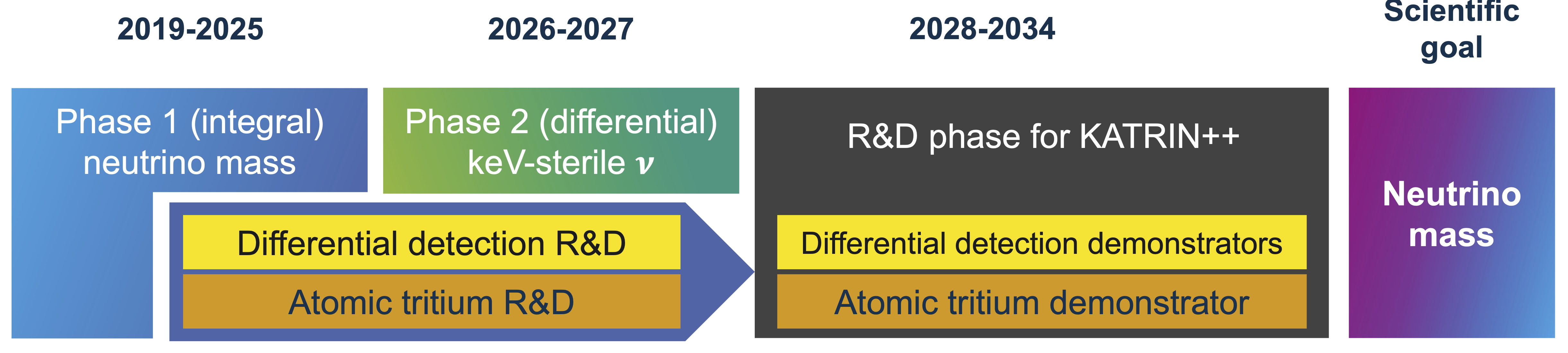}
    \caption{Schematic time line of the staged approach of KATRIN towards an envisioned next-generation experiment, KATRIN++ \cite{KATRIN_input_ESPP2024}. Meanwhile it is agreed that the phase 2 with the keV sterile neutrino search will last up to 2028, copyright Magnus Schlösser.
    \label{fig:KATRIN_time_line}}
\end{figure}

In 2025 the KATRIN experiment is coming to the end of its phase 1 after having collected 1000 measurement days of tritium endpoint data in the search for the neutrino mass. The final neutrino mass sensitivity of KATRIN phase 1 will be below 0.3\,eV (here the sensitivity is defined as the upper limit with 90\,\% C.L. in case no neutrino mass is found).
Fig. \ref{fig:KATRIN_time_line} exhibits the staged approach of KATRIN towards a even more sensitive experiment KATRIN++. 

\begin{figure}[h]
    \centering
    \includegraphics[width=0.38\textwidth]{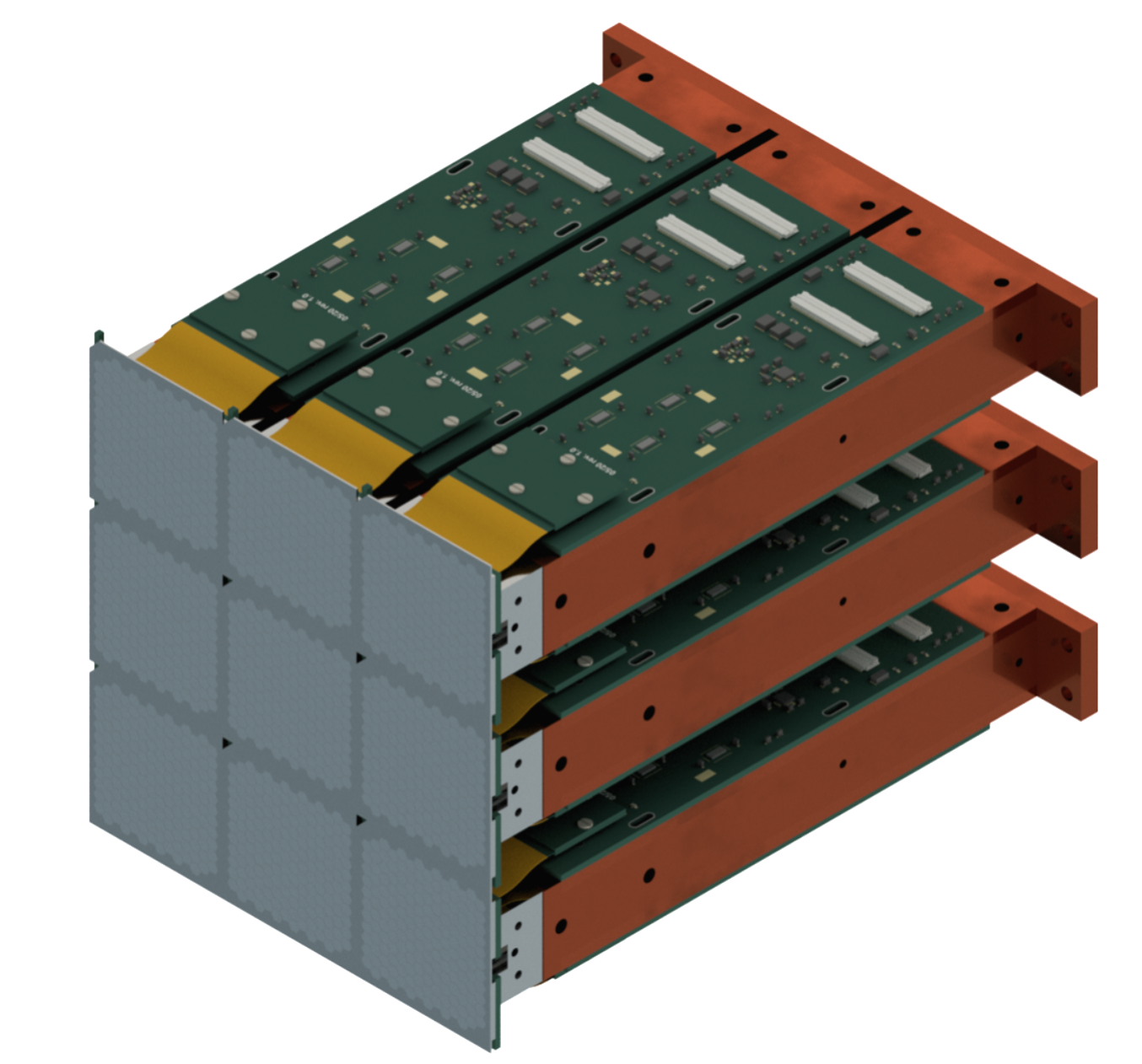}
    \includegraphics[width=0.6\textwidth]{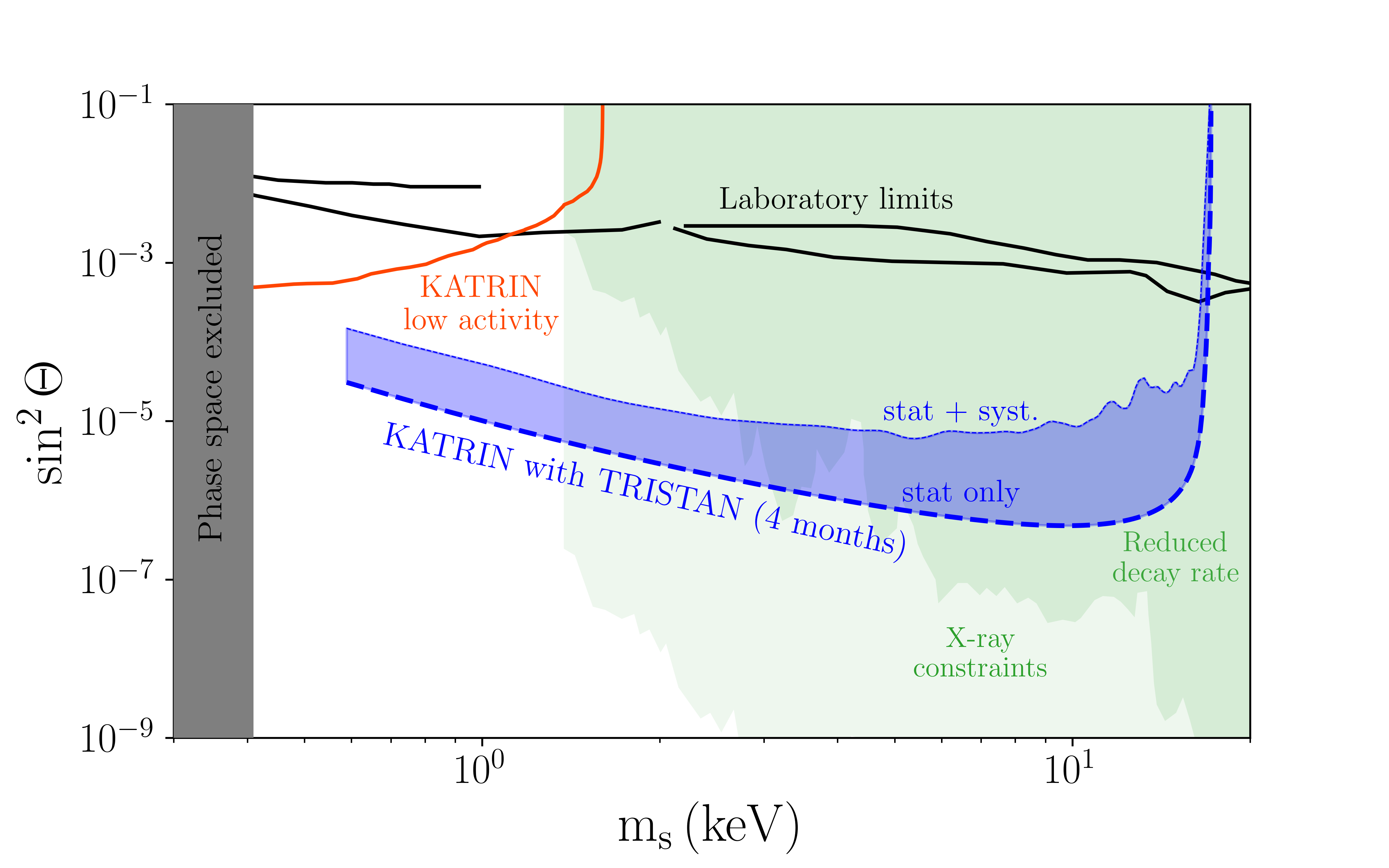}
    \caption{The TRISTAN focal plane detector upgrade for KATRIN's phase 2 searching for sterile keV neutrinos (left) and the projected sensitivity of search in the plane of the fourth mass eigenstate $m_\mathrm{s}$ and the mixing angle to the electron neutrino $\Theta$ (right) \cite{KATRIN_input_ESPP2024}. Here the blue dashed line corresponds to the statistical uncertainty and the blue band above to the possible impact of systematic uncertainties for 4 months of data taking, whereas the light (dark) green area shows the exclusion limits from indirect searches (assuming certain models) 
    The black lines show various previous laboratory limits including the first keV-sterile neutrino search by KATRIN based on an exaploratory first run, \cite{KATRIN:2022spi} and see details therein. Copyright Susanne Mertens and group.
    \label{fig:tristan}}
\end{figure}

For {\bf KATRIN phase 2} the focal plane detector will be upgradeded by the TRISTAN array \cite{Carminati:2023cqf}, see Fig. \ref{fig:tristan}, to investigate the tritium \bspec\ with high precision over the 
broad range $1\,\mathrm{keV} < E < E_0$, primarily to search for keV sterile neutrinos, which are interesting warm dark matter candidates, and to hunt for other beyond the Standard Model signatures.
The signature of a sterile keV neutrino is similar to the signature of an eV sterile neutrino, see Fig. \ref{fig:beta_spec_sterile_eV}, but just extended over a broader energy range. 

The TRISTAN detector consists of 9 silicon drift detector modules with 166 pixels each. These detector modules together with the  front-end electronics are directly mounted on the cooling block. This detector upgrade possesses two major and distinct advantages for investigating the \bspec\ of tritium over nearly the full range:

\begin{itemize}
    \item An essential feature of these silicon drift detectors is that the charge created by the impacting particle is drifted by a radial drift field to a very small central read-out anode with ultra-low capacitance and an embedded FET and nearby read-out electronics. For $\gamma$ radiation this detectors exhibits  energy resolutions close to the Fano limit. This feature together with the very thin dead layer enables energy resolutions of less than 300\,eV (FWHM) for 18.6 keV electrons, which is thus more than a factor 5 better compared to the performance of the standard KATRIN detector. 
    \item The TRISTAN detector with its feature of having more than 10 times more pixels can accept and process much higher electron rates and is thus capable for going deep into the tritium \bspec\ using a low-activity tritium source.
\end{itemize}
With the TRISTAN detector, the \bspec\ will not to be measured in integral mode as in the search for the neutrino mass in the endpoint range, where the count rate at the detector is measured as a function of the retarding energy $qU$. Instead, a fixed and low retarding threshold $qU$ is set so that the energy spectrum of the electrons transmitted above this threshold is determined directly from the high-resolution detector spectra. With several thresholds, the \bspec\ will be measured from approximately 1 keV to \ezero . The signature of a keV-scale neutrino with mass $m_\mathrm{S}$ extends over a broad range of several $m_\mathrm{S}$ (see Fig.\ref{fig:beta_spec_sterile_eV}), so that the energy resolution of the silicon drift detectors of $\Delta E<300$\,eV is more than sufficient for this purpose. The same applies to most of the other Beyond the Standard Model signatures to be investigated. Simulations show that after only a few months measurement time the statistical sensitivity will already reach a few times $10^{-7}$ for the mixing parameter $\sin^2{\Theta}$ statistically and proceeds into the $10^{-6}$ region including systematics (see Fig. \ref{fig:tristan} right). To improve further, the main challenge will be to keep the systematics uncertainties small enough to take advantage of higher statistics.


\begin{figure}[h]
    \centering
    \includegraphics[width=0.6\textwidth]{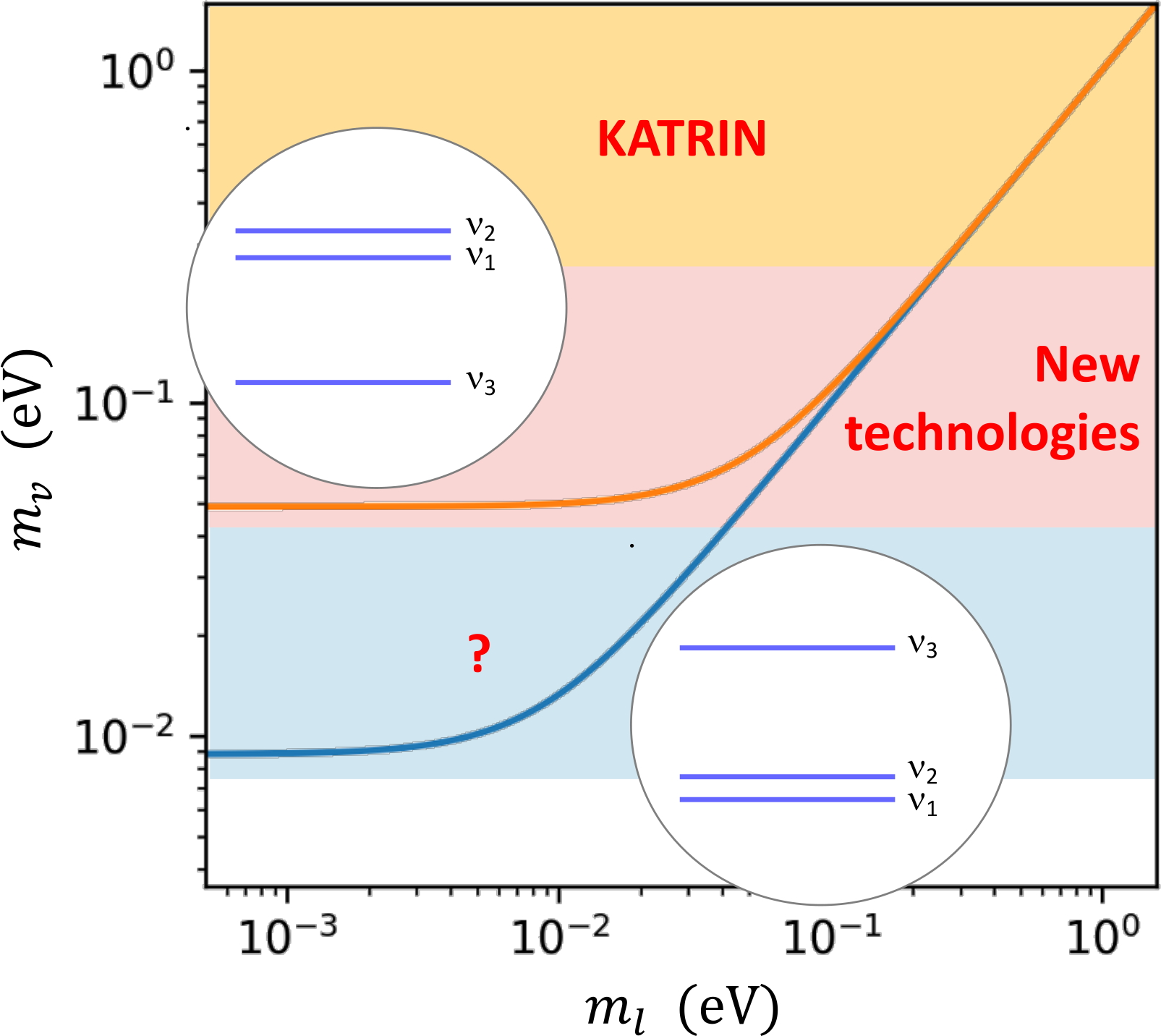}
    \caption{Possible values of the effective electron neutrino mass $m_\nu$ for inverted (orange line) and normal (blue line) ordering as function of the lightest neutrino mass eigenstate $m_l$ (equal to $m_1$ ($m_3$) in the case of normal (inverted) neutrino mass ordering). The KATRIN sensitivity range is illustrated as yellow band. Within the KATRIN++ program the full inverted ordering scheme is to be investigated with the new technologies being developed in the KATRIN++ R\&D phase (pink band), whereas the technology remains yet to be developed to reach
  a neutrino mass sensitivity extending down to 9\,meV  (light blue band).
    This ultimate goal however is being aimed for by the envisioned next-generation experiment KATRIN++, courtesy and copyright Susanne Mertens).
    \label{fig:KATRIN++}}
\end{figure}

Fig. \ref{fig:KATRIN_time_line} illustrates that in parallel to phase 2, dedicated R\&D work for an upgrade to {\bf KATRIN++} will begin. The goal of the KATRIN++ project is to continue using the KATRIN facility to a large extent, but to make use of its full potential via
a \emph{smart} upgrade that is now possible by modern quantum technology. 
The goal of KATRIN++ is thus to measure the effective electron neutrino mass directly in the laboratory, which has become a key challenge of physics due to the puzzles in view of the modern cosmological analyses. In a first step, a neutrino mass sensitivity of 50 meV is to be achieved, which requires an improvement by a factor of 36 in the observable \mnuetwo . As Fig. \ref{fig:KATRIN++} shows, this will allow the effective electron neutrino mass to be measured reliably if the neutrinos have an inverted mass ordering. In order to be able to measure the neutrino mass with certainty even in case of normal mass ordering, a sensitivity of 9 meV must be achieved. While no possible method is yet apparent for the second case, there is a clear idea of how a neutrino mass of down to 50 meV can be measured with KATRIN++ in the first case.

\begin{figure}[t]
    \centering
    \includegraphics[width=0.65\textwidth]{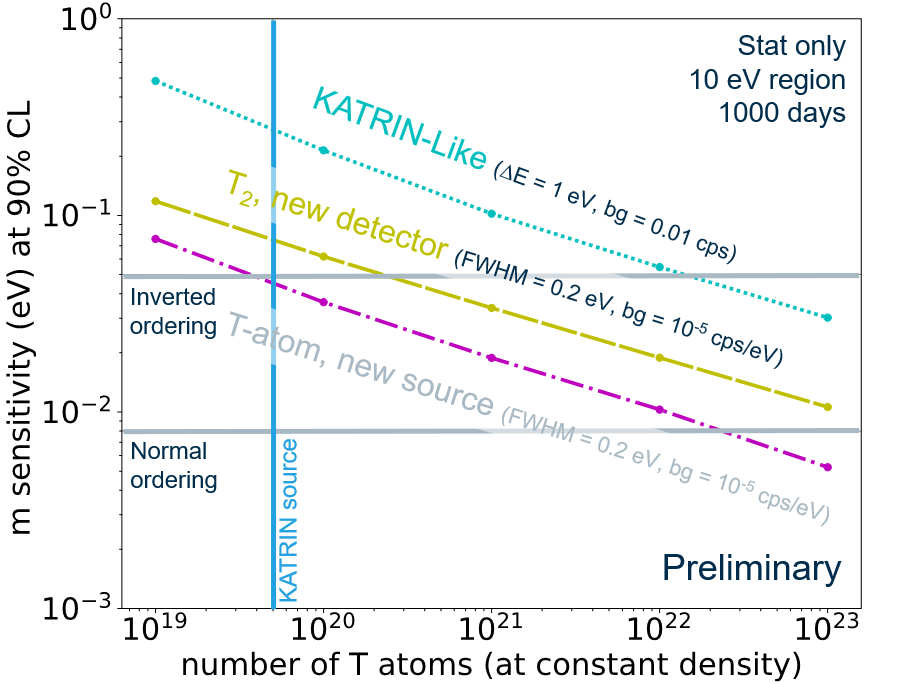}
    \caption{Sensitivity estimate for three different KATRIN/KATRIN++ scenarios as a function of the number of tritium atoms in the source for 1000 days of measurement time over the last 10 eV below the endpoint, without taking systematic uncertainties into account. The KATRIN experiment with molecular tritium, with the design values for energy resolution and background rate (light blue), is clearly surpassed when a differential measurement method with five times better energy resolution and very low background (ochre) can be used. A further improvement is possible if the rotational-vibrational broadening of the final states disappears by switching to an atomic tritium source (magenta) \cite{KATRIN_input_ESPP2024}, copyright Susanne Mertens.
    \label{fig:KATRIN++_sensitivity}}
\end{figure}

The upgrade to KATRIN++ primarily consists of enabling differential measurement of the endpoint range of tritium using quantum technology. Fig. \ref{fig:KATRIN++_sensitivity} shows the improvement in sensitivity achieved by switching from the current integral (light blue) to a new differential method (ochre), which is mainly due to the enormous gain in statistics. While KATRIN measures the endpoint region by sequentially applying about 30 different retarding voltages and measuring the respective count rate, the differential measurement aimed at with KATRIN++ should proceed by setting only one retarding voltage and simultaneously measuring the \bspec\ above this retarding threshold with an energy resolution in the sub-eV range. In addition, such a differential measurement at a fixed retarding potential $qU$ will also reduce the dominant background component that is generated within the volume of the spectrometer and arrives at the detector with an energy very close to $qU$. KATRIN++ is investigating two possibilities for implementing this differential method:
\begin{itemize}
    \item One way to measure this difference is to place a cryogenic metallic microcalorimeter (MMC) array behind the analysis plane of the KATRIN spectrometer. At first glance, this resembles the technology used in the ECHo and HOLMES experiments, but it is a rather different approach. Since only the upper part of the spectrum near its endpoint is transmitted by the retarding voltage of the spectrometer, there are no pile-up problems and much higher statistics can be collected. Furthermore, a clear energy resolution below eV is targeted. Due to the usual incompatibility of cryogenic microcalorimeters with strong magnetic fields, the area of this detector array must be significantly larger than that of the current focal plane detector. In addition, the microcalorimeters must be further developed so that they can keep their superior energy resolution  in magnetic fields of at least several 10 mT. In addition, such a cryogenic microcalorimeter array must be operated in a challenging temperature environment of only a few tens of millikelvin, i.e. it must be protected from radiation at room temperature by a cryogenic aperture.
\item Another way to measure the spectrum of the transmitted \belec s simultaneously while setting only one retarding voltage is time-of-flight spectroscopy \cite{Steinbrink:2013ska}. It is based on the fact that the velocity of electrons just above the threshold of the retarding potential is small and thus strongly depends on the amount of excess energy above the threshold. This method is frequently used at KATRIN for calibration measurements with  the photoelectron source irradiated with a pulsed UV laser \cite{KATRIN:2021rqj}. To successfully apply this method to tritium $\beta$ spectroscopy would require the development of a minimally invasive tagger for electrons that  determines their entering of the spectrometer with a time resolution of better than 100 ns.
There are several approaches to realise such a device, all of which have in common a highly sensitive readout of very weak signals, again relying on quantum technology at low temperatures.
Secondly, the energy resolution of the spectrometer must be significantly improved to below 1 eV, which can be achieved not only by using a larger spectrometer but also with the help of the novel method labelled the transverse energy compensator (TEC) \cite{KATRIN_input_ESPP2024}, which performs manipulation in phase space using a rapidly decreasing electrical potential on a drift tube in front of the main spectrometer.
\end{itemize}
Demonstrators for the two differential methods are to be built from 2028 onwards in order to down select the final technology.
Fig. \ref{fig:KATRIN++_sensitivity} shows, however, that the two differential methods mentioned, with their energy resolution in the sub-eV range (ochre), will not be sufficient on their own to achieve a sensitivity of 50 meV unless the number of tritium atoms in the source is drastically increased. This is because the rotation-vibration-induced broadening of the final states of molecular tritium will lead to an effective blurring of the energy resolution with a nearly Gaussian width of 400 meV \cite{Schneidewind:2023xmj}. Therefore, a further independent R\&D line at KATRIN++ is focused on the development of an atomic tritium source (magenta), which is being carried out in close collaboration with the Project 8 collaboration which persues similar goals.

Due to the need to first find a suitable technology that can achieve a significant improvement in neutrino mass sensitivity, and the many projects listed in Sec. \ref{compl} for the direct determination of the effective electron neutrino mass, some efforts will certainly be postponed in favour of others and projects will be merged in the future.

\clearpage 

\section{Conclusions}
\label{sec:conclusions}
The fundamental and far-reaching question of the neutrino mass scale, which was put to the forefront after the discovery of neutrino oscillation, has still not been completely resolved, but the KATRIN experiment located at Karlsruhe Institute of Technology (KIT) has significantly narrowed down the unexplored parameter space. KATRIN measures the endpoint region of the tritium spectrum with unprecedented accuracy, combining a windowless gaseous molecular tritium source with $10^{11}$ \bdec s per second with a very high-resolution, integrating spectrometer with large solid angle acceptance. Another asset of KATRIN is its many calibration and diagnostic instruments and methods that have resulted in extremely small systematic uncertainties. The measurement data evaluated so far give an upper limit on the effective electron neutrino mass of 0.45\,eV (90\,\% C.L.). The final sensitivity of KATRIN, based on six times more measurement data, will be below 0.3\,eV. The KATRIN Collaboration has
also successfully analyzed its set of 
 precise  data from  tritium \bdec\ near its endpoint for many other searches for physics beyond the Standard Model, in particular for sterile neutrinos with masses in the eV range.

KATRIN's phase 2, which will begin in 2026, is dedicated to measuring nearly the entire spectrum of tritium with unprecendented precision, in particular to search for sterile keV neutrinos, an interesting candidate for warm dark matter. 

However, KATRIN's excellent apparatus at KIT will continue to be used in the future 
to push for the ultimate sensitivity in measuring the effective electron neutrino mass. Under the label KATRIN++, an R\&D programme is currently being launched to improve the neutrino mass sensitivity down to 50\,meV in a first step through two decisive measures: a differential read-out method based on modern quantum technology in combination with an atomic tritium source, which together will significantly reduce statistical and systematic uncertainties. 

Will this enable the neutrino mass to be measured reliably? Yes, at least in case that the neutrino mass ordering is inverted. However, if the ´ghost particles´  of the universe would follow the normal mass ordering, which we will learn within the next years by neutrino oscillation experiments such as JUNO, Hyper Kamiokande, DUNE and ORCA, then in the worst case and preferred by todays cosmological analyses, a sensitivity of 9\,meV must be achieved in the direct determination of neutrino masses in the laboratory.
 The technology for this is not yet fully  conceptualized, but the KATRIN++ developments and other projects in this and the wider scientific community will certainly make it possible to achieve this most important step over the coming decades.

\section*{Acknowledgments}
We would like to express our heartfelt thanks to the outstanding, highly committed and extremely collaborative environment provided by our current and former members of the international KATRIN collaboration, to Karlsruhe Institute of Technology (KIT) for its generous hospitality, commitment and leading role in this challenging project. Our special thanks go to the colleagues Alexey Lokhov, Susanne Mertens, Magnus Schlösser, Jaroslav \v{S}torek, Kathrin Valerius, Michael Zacher who provided special figures to our article.\\
We would like to express our sincere gratitude to the various funding institutions, in particular the German Federal Ministry of Research, Technology and Space (BMFTR), the Helmholtz Association, the German Research Foundation (DFG), the U.S. Department of Energy, the Czech Academy of Sciences, as well as to the highly supportive international scientific community, which made this fantastic KATRIN project a reality.

\clearpage 
\bibliographystyle{unsrt}
\bibliography{reference4}

\end{document}